\documentclass[aps,amsmath,prb,twocolumn]{revtex4}
\usepackage{amssymb}
\usepackage{amsbsy}
\usepackage{amsmath}
\usepackage{float}
\usepackage[pdftex]{graphics}
\usepackage{graphics}
\usepackage{color}
\usepackage{graphicx}
\usepackage{subfigure}
\usepackage{bm}
\usepackage{hyperref}
\usepackage{natbib}
\newcommand{\comm}[1] {\textcolor{black}{#1}}

\def\be{\begin{equation}}
\def\ee{\end{equation}}

\begin{document}

\title{Onset of Floquet thermalisation}

\author{Asmi Haldar$^{1}$, Roderich Moessner$^2$ and Arnab Das$^1$}

\affiliation{$^1$Department of Theoretical physics, Indian Association for the Cultivation of Science, \\
2A \& 2B Raja S. C. Mullick Road, Kolkata 700032, India} 

\affiliation{$^2$Max Planck Institute for the Physics of Complex Systems, \\
	N\"{o}thnitzer Stra{\ss}e 38,
01187 Dresden, Germany }


\begin{abstract}
In presence of  interactions,  a closed, homogeneous (disorder-free)
many-body system is believed to generically heat up to an `infinite temperature' ensemble 
when subjected to a periodic drive: in the spirit of the ergodicity hypothesis
underpinning statistical mechanics, this happens as no energy or other conservation
law prevents this. Here we present an interacting Ising chain driven by a  field of time-dependent strength, 
where 
such  heating onsets only below a threshold 
value of the drive amplitude, above which the system exhibits non-ergodic behaviour.
The onset appears at {\it strong, but not fast} driving. This in particular puts
 it beyond the scope of high-frequency
expansions. The onset
location shifts, but it is robustly present, across wide variations of the model Hamiltonian
such as driving frequency and protocol, as well as the initial state.  The portion of nonergodic
states in the Floquet spectrum, while thermodynamically subdominant, has a finite entropy. We find that 
the magnetisation as an {\it emergent} conserved quantity underpinning the freezing; indeed
the freezing effect is readily observed, as initially magnetised states remain partially
frozen {\it up to infinite time}.
This result, which bears a family resemblance to the Kolmogorov-Arnold-Moser theorem for classical dynamical systems,
could be a valuable ingredient for extending Floquet engineering to the interacting realm.
\end{abstract}
    
\maketitle
\section{Introduction}
Interacting many-body systems, by the ergodic hypothesis, generically thermalise, placing them 
in the purview of statistical mechanics and equilibrium thermodynamics~\cite{Huang_Stat_Mech}. 
Our understanding of the corresponding situation for non-equilibrium systems is still in flux. For perhaps 
the simplest class
of non-equilibrium systems, periodically driven (Floquet) systems, thermalisation physics 
at first pass looks maximally simple: removing time translation invariance destroys energy conservation, 
and hence the concept of temperature--which means  thermalisation is to a featureless `infinite-temperature' 
state~\cite{LDM_PRE, Rigol_Infinite_T}. 

Such Floquet systems have been predicted to be  capable of sustaining new forms of spatiotemporal ordering 
when many-body localised as a result of strong quenched disorder~\cite{KLMS}. The experimental search for 
such so-called discrete time crystals has been qualitatively more successful~\cite{DTCions,DTCnvcentres} 
than may have been anticipated: the collection of systems appearing to exhibit such order now even includes 
a dense periodic array of nuclear spins initialised in a thermal 
state~\cite{barrett_NMR_DTC}. 

All of this focuses the question on settings which permit long-lived correlations and order to persist despite 
the presence of periodic driving even in the absence of quenched disoroder. 
In periodically driven {\it non-interacting} systems, quantum 
heating can be suppressed~\cite{AD-DMF,Mahesh_Freezing,Russomanno_JStatMech} and an
extensive number of periodically conserved quantities identified~\cite{PGE}. 
In turn, a pre-thermalisation regime has been
identified~\cite{Eckstein_Canovi} which resembles a frozen non-thermal state
~\cite{AD-DMF} which can be described by a 
periodic (generalized) Gibbs' ensemble~\cite{PGE}.  Tuning the drive parameters, and weakening the interactions, 
can substantially enhance the prethermalization period, still expected to remain 
finite~\cite{Dima_Floquet_Prethermalization}. In fact, 
for disorder-free systems, a transient but exponentially long-lived
regime exhibiting discrete time-crystalline phenomenology
has already been identified~\cite{else_nayak_prethermal}.
These constitute lower bounds on the thermalisation timescales. 
For finite-size systems, an emergent integrability structure
for strong drives has also been proposed as a way to avoid thermalisation~\cite{gil-fss}.
There is further evidence indicating  absence of heating at high drive frequencies in a 
variety of other settings~\cite{Kris-Periodic,Luca_Polku,
Bukov_Polku_Huse,Anatoli_Rev,Adhip_Diptiman,Sayak_Utsa_Amit, Bordia_Knap_Bloch,Sreejith_Mahesh,
Qin_Hofstetter} and in specially designed models~\cite{Prosen_prl_98,AL_DL_PRB}.

Here, we address the question whether there is an identifiable threshold for the ratio of 
driving and interaction strength,
 below which the system approaches a 
non-trivial steady state that depends on the drive and the initial state. We consider a  spin chain subject to {\it strong, but not fast}
driving, and use remanent infinite-time magnetisation of a  initial magnetised state as measure of failure to Floquet-thermalise. 
As the driving is increased from low strength, where standard Floquet thermalisation is observed, we find a remarkably
well-defined second regime, in which remanent magnetisation is present 
even in the infinite time limit. Its value is given by the Floquet diagonal ensemble average implied by the initial state. 
The location of this threshold moves, but its existence is stable to variations in state initialisation, 
driving strength, driving protocol, and driving frequency. 

In all cases, however, we are able to identify an emergent {\it approximately} conserved quantity -- in the case we discuss at length, 
the magnetisation itself -- which
becomes exactly conserved if the static part of the Hamiltonian is ignored. Thus, rather than an extensive set of integrals of motion,
as is present in the case of the periodic Gibbs ensemble\cite{PGE} and the Floquet many-body localised 
cases\cite{FlqMBL1,FlqMBL2,FlqMBL3}, all that appears to be needed
to stop the system from heating up indefinitely is a single, approximately conserved quantity. 

While our numerical investigation on systems up to $L=14$ spins naturally limits our capacity to extrapolate these results to the `thermodynamic' limit,
there are indications that this is not only a finite-size effect. Firstly, in plots of remanent magnetisation versus driving strength, 
we identify a crossing point for curves for different $L$ separating the ergodic and the non-thermal regimes. 
Second, the set of Floquet eigenstates exhibiting memory, while accounting only for
 a vanishing fraction of the total Hilbert space, extrapolates to have 
 a finite entropy in the thermodynamic limit. This means 
 that such states can still be straightforwardly selected by an initial condition, not
unlike initialising a static system in a low-temperature configuration. 

In the following, we first we set the notation and provide a brief introduction to the 
Floquet concepts we have used. We then define our model and  drive protocol. 
We characterise the ergodic and the non-thermal phases and the threshold
between them using various measures, and demonstrate robustness to variations of drive patterns and system parameters. 
We close with an outlook and suggestions for further investigations. In particular, 
origin and nature of the sharp features in the memory as a function of driving strength merit further study. 


\section{Model}
In this section, we introduce notation, model Hamiltonian, drive protocol, and observables to be studied.

\subsection{Floquet basics}	
Let us decompose the time-dependent Hamiltonian $H(t)$ into 
 a static interacting Hamiltonian $H_{0}$ and a time-periodic drive
$H_{D}(t)$ with $[H_{0},H_{D}] \ne 0$: 
\begin{equation}
H(t) = H_{0} + H_{D}(t),\label{eq:H0-hd}
\end{equation}
\noindent
The time evolution operator evolving a state through a period from
$t=\epsilon$ to $t=\epsilon + T$ ($0  \le \epsilon < T$) is $U (\epsilon)$. 
Since $U(\epsilon)$ is unitary, it can always be
expressed in terms of a hermitian operator, the `Floquet Hamiltonian' $H_{eff}$ as
\begin{equation}
U(\epsilon) = e^{-iH_{eff}(\epsilon)T}. 
\label{Heff_def}
\end{equation}
Formally, 
\begin{equation}
\exp\left(-iH_{eff}\left(\epsilon\right)T\right)
= \mathcal{T}\exp\left(-i\int_{\epsilon}^{\epsilon+T}dt\: H(t)\right),\label{eq:defn-heff}
\end{equation}
\noindent where ${\mathcal T}$ denotes time-ordering. 
Let $|\mu_{i}\rangle$ denote the $i$-th `Floquet eigenstate' of 
$H_{eff}$ corresponding to the `Floquet eigenvalue'  (also known as quasienergy) $\mu_{i}$.

A sequence of stroboscopic observations at instants 
$t = \epsilon, \epsilon + T, \dots, \epsilon + nT$ (integer $n$) is 
identical to that produced by the dynamics under the time-independent Hamiltonian 
$H_{eff}.$ 
This applies for every $\epsilon$,  
hence we get continuous family of stroboscopic series. 

In the following,
we are interested in long-time asymptotic behaviour, so that temporal variations within 
a driving period are of secondary importance.  Hence, we arbitrarily pick $\epsilon = 0.$ 

\subsection{Infinite time limit: diagonal ensemble average}

The nature of the asymptotic state under the drive can be understood as follows. 
Consider an initial state
$$|\psi(0)\rangle = \sum_{i}c_{i}|\mu_{i}\rangle$$ 
\noindent
and the stroboscopic time series for an observable 
$$\hat{\cal{O}} 
= \sum_{i,j}{\cal O}_{ij}|\mu_{i} 
\rangle\langle \mu_j|.$$
\noindent
\begin{equation}
	\langle \psi(nT+\epsilon) |{\hat {\cal O}}|\psi(nT+\epsilon) \rangle 
	= \sum_{i,j}c_{i}c_{j}^{\ast}{\cal O}_{ij}e^{-i(\mu_{i}-\mu_{j})(nT+\epsilon)}. 
\label{O_exp}
\end{equation}
\noindent
Like in the case of static Hamiltonians, 
under quite general and experimentally relevant conditions (see, e.g., Ref.~\cite{Reimann}), 
at long times ($n\to\infty$) the off-diagonal ($i\ne j$) 
terms `average to zero' and 
the state of the system can hence be described by an effective ``diagonal ensemble" (in the absence of synchronisation, e.g.
for discrete time crystals, this is replaced by a block diagonal ensemble\cite{Lazarides_RM_BDE}). 
This is captured by the mixed density matrix~\cite{Rigol_Nature}
$
{\hat \rho}_{_{DE}} = \sum_{i}|c_i|^2 |\mu_{i}\rangle\langle\mu_{i}|.
$

Thus, the asymptotic properties of a periodically driven system are 
effectively given by a classical average (known as diagonal ensemble average or DEA) over the expectation 
values of the eigenstates of $H_{eff}$,
\begin{equation}
\langle \hat{\cal O}\rangle{\rm (DEA)} = \sum_{i}|c_i|^2 \langle\mu_{i}|\hat{\cal O}|\mu_{i}\rangle \ .
\label{eq:odea}
\end{equation}
Hence it is sufficient to study the nature of the eigenstates and eigenvalues of $H_{eff}$, or equivalently of $U(\epsilon)$, 
in order to obtain the long-time behaviour. 

\subsection{Driving protocol}
We consider  $L$ spins on a chain. We chose a binary drive protocol, which switches periodically between a pair of rectangular pulses. 
The time dependent Hamiltonian is 
\begin{equation}
H(t) = H_{0} + \mathrm{sgn}(\cos{\omega t})\, H_{D}, 
\end{equation}
with the two components
\begin{eqnarray}
H_{0} &=& -J\sum_{i}\sigma_{i}^{x}\sigma_{i+1}^{x} +\kappa\sum_{i}\sigma_{i}^{x}\sigma_{i+2}^{x} -h_{0}^{x}\sum_{i}^{L}\sigma_{i}^{x}
\nonumber \\ 
&&- h^{z}\sum_{i}^{L}\sigma_{i}^{z}; \\
H_{D} &=& - h_{D}^{x}\sum_{i}^{L}\sigma_{i}^{x}\ . 
\end{eqnarray}
The $\sigma^\alpha$ are Pauli matrices. We use periodic boundary condition, but tamper the system  
slightly by putting $J_{L,1} = 1.2J$ and $\kappa_{L-1,1}=1.2\kappa$ to break translational invariance (and 
hence remove any remaining block-diagonal structure of the Hamiltonian). Here since we keep the interaction strengths
constant during the drive, we use the drive amplitude itself 
as the tuning parameter.

In presence of the transverse field, the Hamiltonian $H_{0}$ is known to be ergodic due to the four-fermionic
interaction terms arising from the next-nearest neighbour interactions under the spin to fermion mapping, 
and also due to the longitudinal field. We  have explicitly verified that $H_{0}$ is ergodic for our case, see Suppl.\ Mat.). 

We initialise the simulation in the time domain with different initial states. Unless otherwise stated, we use the default choice
of the ground state of $H(t=0)$. 

\section{Numerical results}
The central quantity is the longitudinal magnetization
\begin{equation}
m^x(t) = \frac{1}{L}\sum_{i}^{L}\langle\psi(t)|\sigma^{x}_{i}|\psi(t)\rangle \ .
\end{equation}
We monitor its real-time dynamics
in a stroboscopic time series. We diagnose non-thermalisation/freezing via 
its long-time asymptotic behaviour, the remnant magnetisation, which we study 
as a function of various model parameters.

\subsection{Onset of Floquet thermalisation}
In the following, we provide numerical evidence that for a strong (but not fast) drive, the system fails to Floquet thermalise, instead retaining
memory of its initially magnetised state. We then show that the onset of Floquet thermalisation occurs at a fairly well-defined threshold
driving strength. For the results in the main text, we have chosen $J=1,$ $\kappa = 0.7,$ $h^{z}=1.2,$ and 
$h_{0}^{x}=0.02.$ 

\begin{figure*}[ht]
\centering
\includegraphics[width=0.32\linewidth]{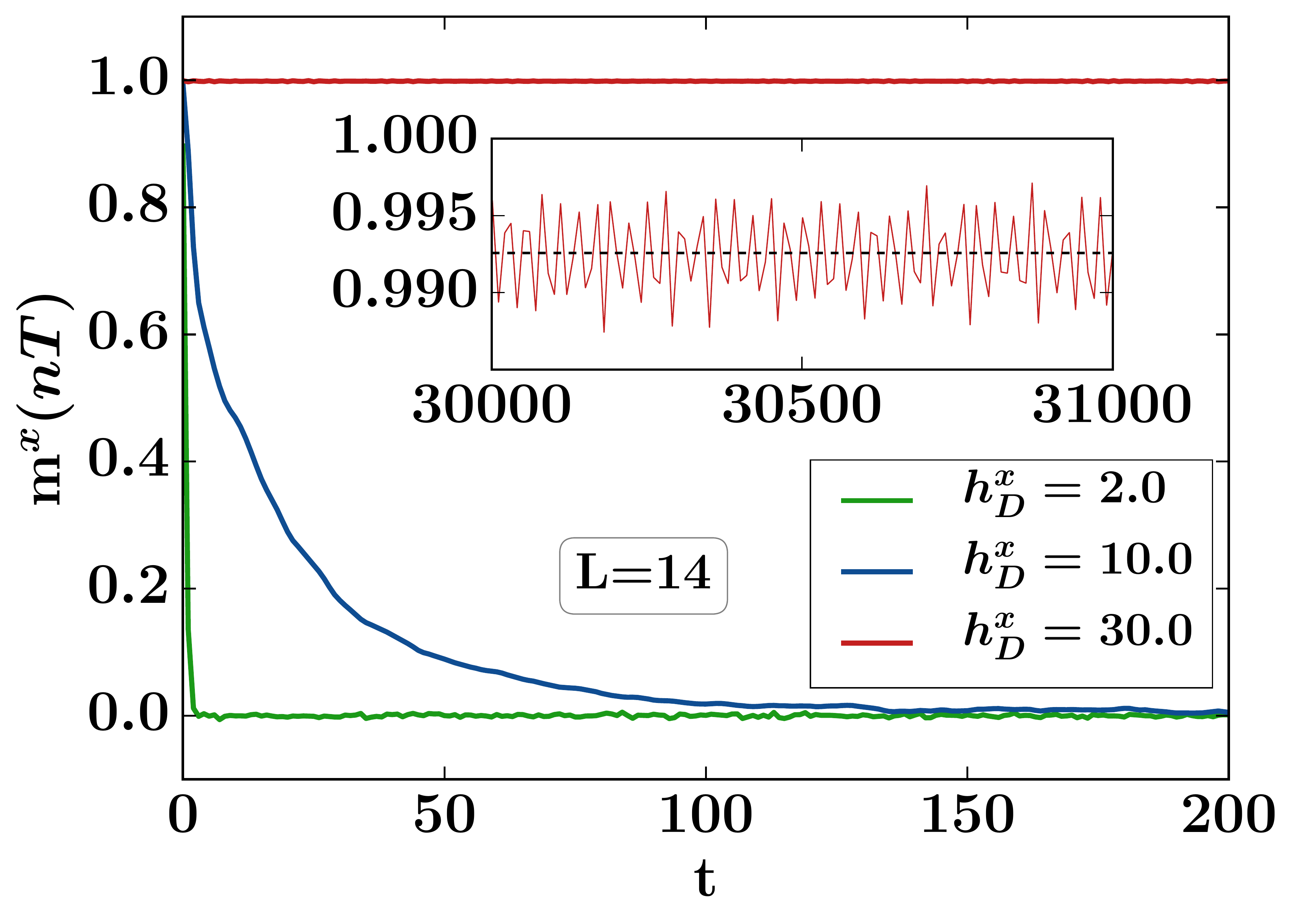}
\includegraphics[width=0.32\linewidth]{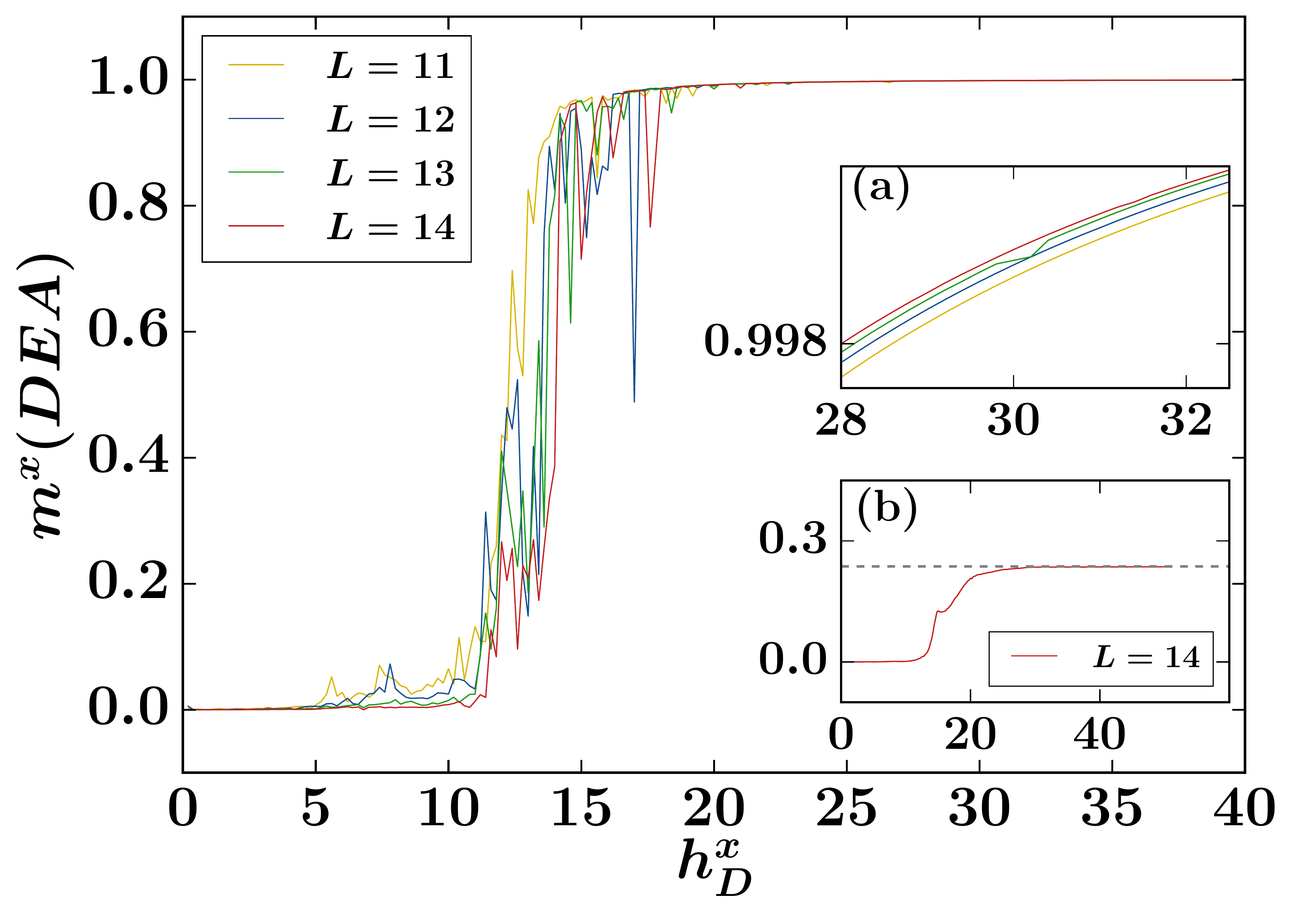}
\includegraphics[width=0.32\linewidth]{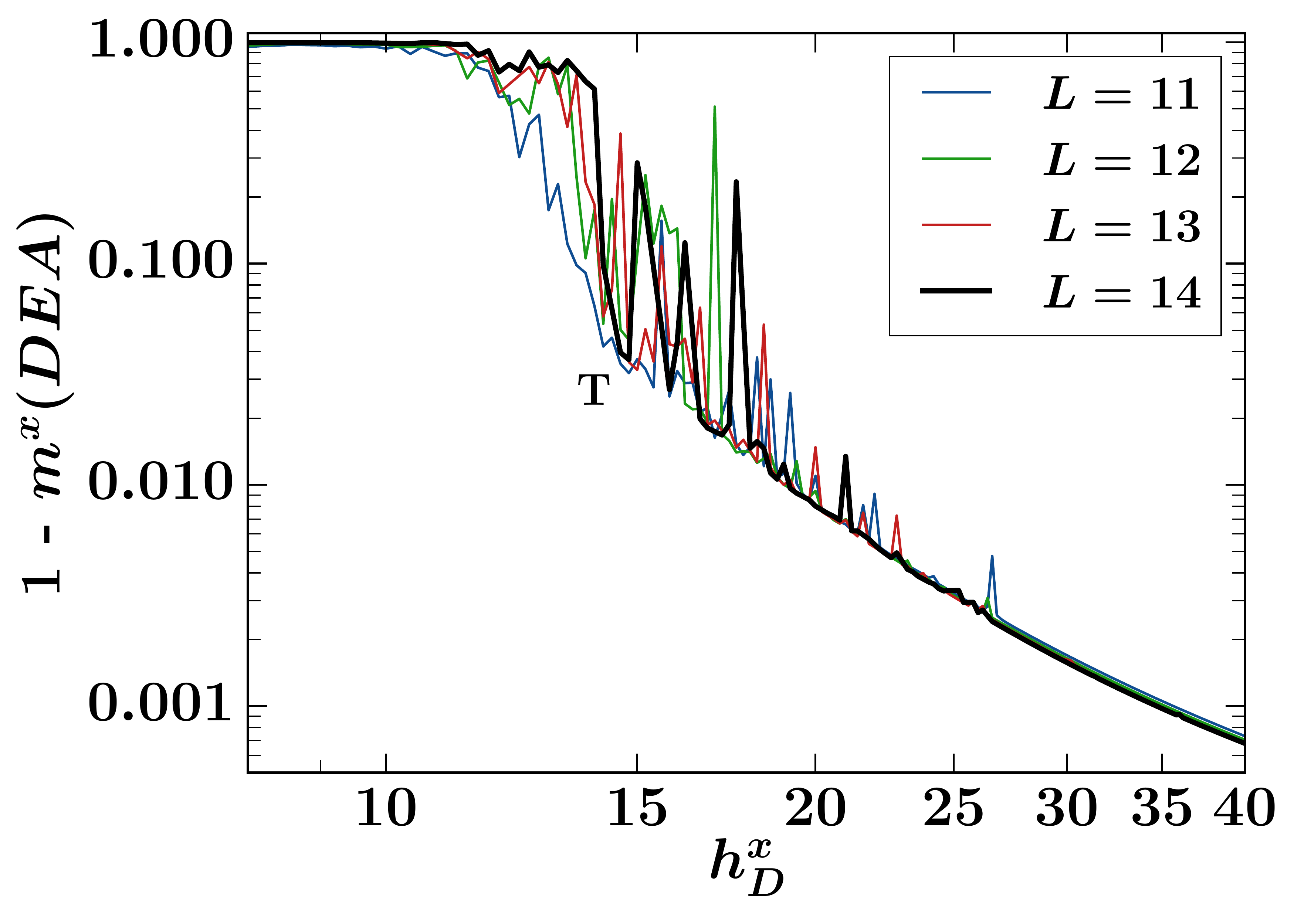}
\caption{
Freezing and its onset threshold.
{\it Left frame}: Stroboscopic time series of magnetisation $m^x(t)$ for different driving
strengths showing initial state memory for strong driving. The inset zooms in on the long-time behaviour; the black 
horizontal line denotes the DEA of the magnetisation. 
{\it Middle frame}: Remnant magnetisation
as a function of driving strength for different system sizes. The 
high-field regime (top inset) shows an {\it increase} of the remnant magnetisation with $L$. The bottom
inset shows the DEA of $m^x$ vs.\  drive amplitude for a `generic' state (see the text for details) whose net 
initial magnetization is marked with the horizontal line, which remains almost unchanged for very strong drives. 
{\it Right frame}: Same data as middle frame on a doubly logarithmic plot for $1 - m^{x}{\rm (DEA)}$ .  The deviation away from 
almost complete thermalisation gets steeper and moves towards the right with increasing system size. The curves
appear to accumulate from the left at a `threshold point' ({\bf T}) which itself appears to move little as the system size
is increased from $L=11$ to $L=14$.
}
\label{fig1:Dynamics_mxDEA}
\end{figure*}

The stroboscopic time series for the magnetisation $m^x$ is shown in Fig.~\ref{fig1:Dynamics_mxDEA}, left frame. Already at short times,
three representative trajectories for different driving strengths show strikingly different behaviour. While for weak driving fields, the 
magnetisation disappears almost immediately, for stronger ones, the decay slows down. Finally, for $h^x_D$ beyond a threshold
value, the decay is arrested: even at the longest times, a remnant magnetisation persists. 

This remnant magnetisation  agrees with  the DEA of the magnetisation evaluated for the same system (see inset). Note that the
nonvanishing DEA is already in itself a signature of the lack of Floquet thermalisation -- in general, Floquet thermalised eigenstates  
individually show no non-trivial correlations. 
	
In order to locate the onset, the DEA of $m^{x}$ as a function of the drive 
amplitude $h_{D}^{x}$  is plotted in Fig.~\ref{fig1:Dynamics_mxDEA}, middle frame. 
	A  threshold for nonzero remnant magnetisation is observed, separating 
	 the ergodic ($m^{x}_{DE} \approx 0$) from the nonergodic regime. 
	 	
The lower inset shows freezing for an initial state with a reduced polarization in the $x-$direction. 
	The black dotted
	line shows the initial value of $m^x$ for the state, and the curve shows that for 
	high enough $h_{D}^{x}$, the DEA of $m^x$ almost coincides with it. 
	In detail, this initial state is given as  $|\psi_0\rangle = \sum_{i=1}^{2^{L}} c_{i} |i_{x}\rangle$, 
	where $|i_{x}\rangle$ is the $i-$th eigenstate of the 
	longitudinal field part (computational basis states in $x$-direction, or $x-$basis states), 
	by choosing $Re[c_{i}]$ and $Im[c_{i}]$ from a 
	uniform distribution between -1 and +1, 
	multiplying them by $e^{\beta m_{i}^{x}},$ where $\beta > 0$ and $m_{x}^{i}$ is the 
	longitudinal magnetization 
	of $|i_{x}\rangle$, and finally normalizing the state. 
	This gives a `generic' state with a bias towards positive longitudinal magnetization. For the plot in
	Fig.~\ref{fig1:Dynamics_mxDEA} (middle frame), we have chosen a random instance corresponding to 
	$\beta = 1.75.$
	The right frame of Fig.~\ref{fig1:Dynamics_mxDEA} shows DEA of $1-m^{x}$ on a 
	doubly logarithmic $log-log$ plot zoomed
	in around the threshold for 
	better visibility. 
	 
\subsection{Floquet eigenstates and an emergent conservation law}

\subsubsection{Localization and magnetization}
We now turn to the properties of the Floquet eigenstates obtained 
by numerically diagonalizing the time evolution operator $U(0)$, Eq.~\ref{Heff_def}.
We consider first their `localisation' in Hilbert space, followed by their magnetisation content.
\begin{figure*}
\centering
\includegraphics[width=0.32\linewidth]{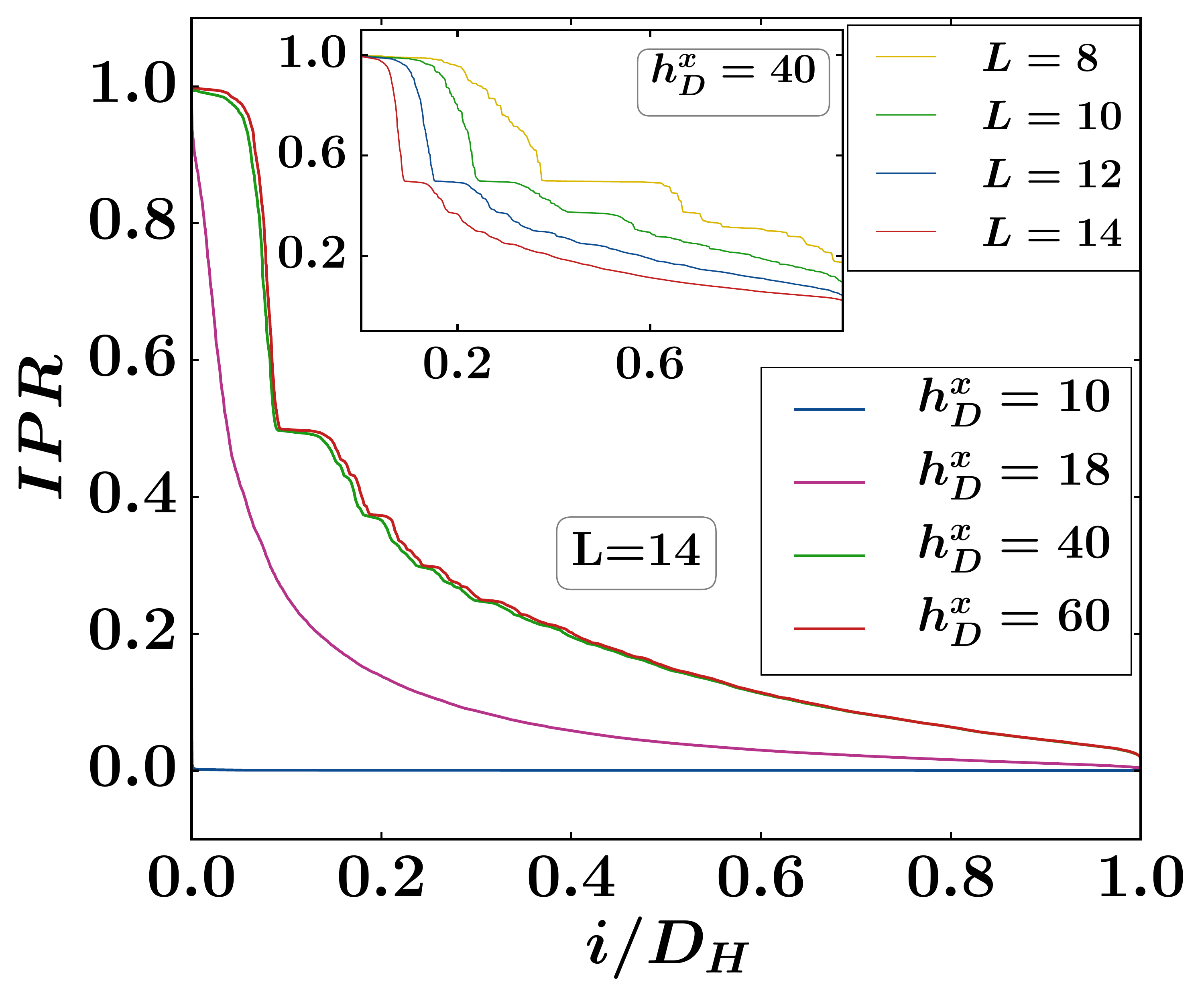}
\includegraphics[width=0.32\linewidth]{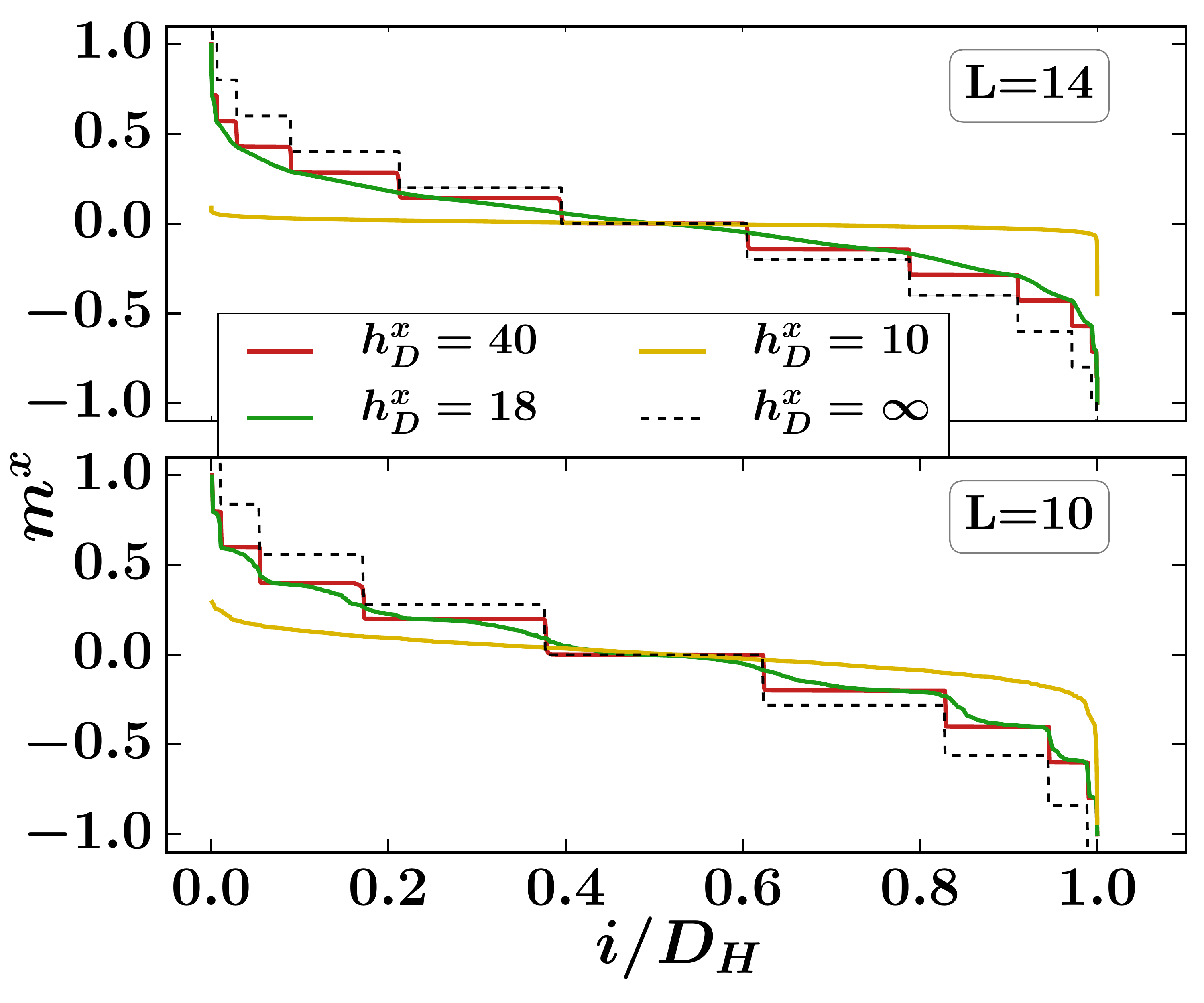}
\includegraphics[width=0.32\linewidth]{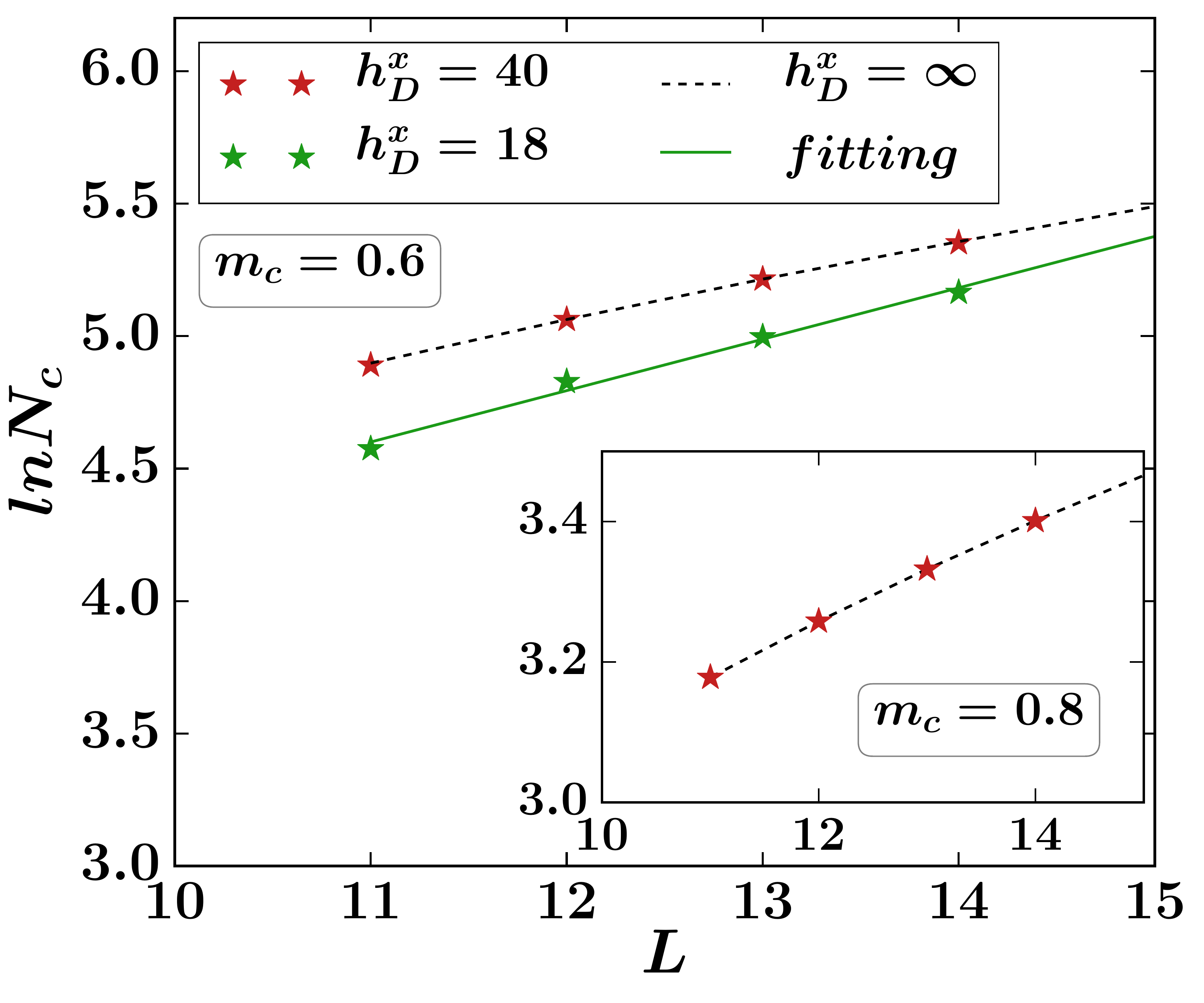}
\caption{
{Emergent conservation law for strong drives, as reflected in the Floquet eigenstates $|\mu_i\rangle$.}
	{\it Left frame}: Values of the IPR in the $x$-basis, arranged in decreasing order. 
	Unbounded heating requires these states to be delocalized in the eigenbasis of any local operator. 
	This is the case for the drive with amplitude 
	below the threshold ($h_{D}^{x} = 10,$) {\it but not above} ($h_{D}^{x} = 18, 40, 60$). 
	The inset shows a decreasing IPR for different system-sizes for $h_{D}^{x}=40$ 
	due to the emergent conservation
	law evidenced in the 
	{\it middle frame:}  $m^x$ for the Floquet eigenstates  arranged in 
	decreasing order, for different values of $h_{D}^{x}$. 
	Black dotted lines  ($h_{D}^{x} = \infty$) show the values of $m^x$ 
	of the $x-$basis states (multiplied by a factor of 1.4 for visibility). 
	For $h_{D}^{x} = 40,$ clear step-like structures appear, indistinguishable from the 
	steps of $m^x$ for $x-$basis states for both system sizes $L=10,14$ (see Suppl. Mat. for finer details
	of $L$ dependence of this matching). For a lower drive value $h^{x}_{D}=18$, close to the threshold, 
	the curve smoothes out, indicating weakening of the quasi-conservation, yet highly polarized Floquet 
	states are still substantial in number. For still 
	lower values (e.g. $h^{x}_{D} = 10$), the curve finally flattens. The pronounced asymmetry in the Floquet 
	magnetizations for lower values of $h_{D}^{x}$ is due to the small asymmetry in the drive. 
	{\it Right frame:}  The log of the number $N_{c}$ of Floquet eigenstates with polarization above a given 
	value $m_{c}$ is shown to  grows approximately exponentially with system size, corresponding to (a 
	vanishing fraction of states but with a) finite entropy. For large $h_{D}^{x}$ ($h_{D}^{x}=40$), the 
	numerical data points fall almost exactly on the analytically calculated (black dotted lines) corresponding
	to $h_{D}^{x}=\infty$ (see the matching of the step-like structures in the middle frame). For a lower value 
	$h_{D}^{x}=18$ a linear fit is done for the numerical data points.
	}
\label{Flq_IPR}
\end{figure*}

In order to investigate the  localization properties of the Floquet states in the $x-$basis
$\{|i_{x}\rangle\}$ we calculate the inverse participation ratio (IPR) in said basis defined as
$IPR(|\mu_{j}\rangle) = \sum_{i=1}^{2^L} |\langle i_{x} | \mu_{j}\rangle|^4$.
The left frame of Fig.~\ref{Flq_IPR} shows the IPR thus obtained, arranged in decreasing order. 
Indefinite heating corresponds to the states being
delocalized in the eigenbasis of any local operator, which implies a uniformly 
small IPR given by the inverse dimension of Hilbert space, $1/D_H$. This is indeed what is observed for small drive 
fields. By contrast, for large drive fields, states appear which have an IPR close to 1, which indicates the presence of well-localised states,
and hence the absence of Floquet thermalisation for the corresponding part of the spectrum. 

\begin{figure*}
\centering
\includegraphics[width=0.35\linewidth]{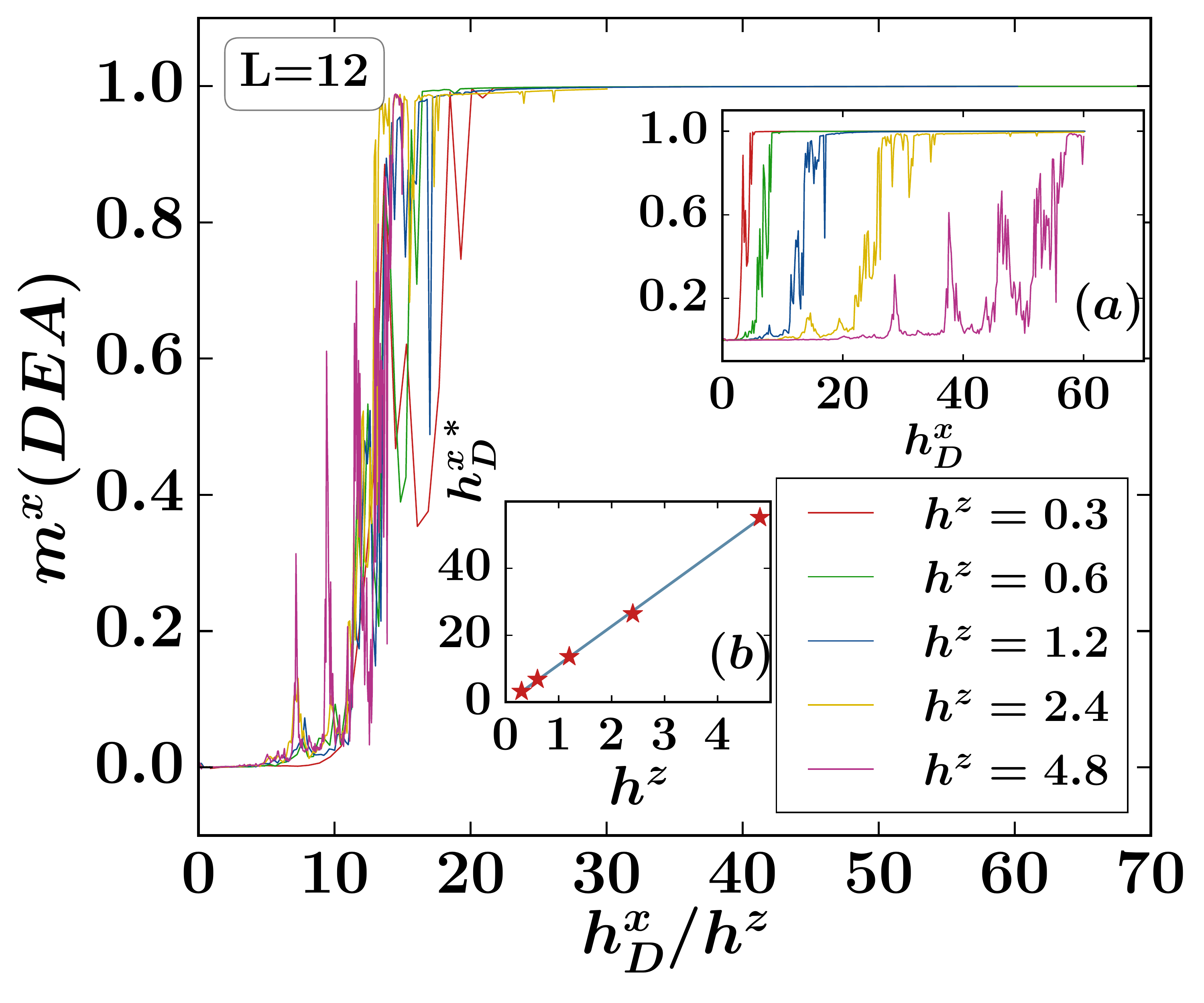}
\includegraphics[width=0.35\linewidth]{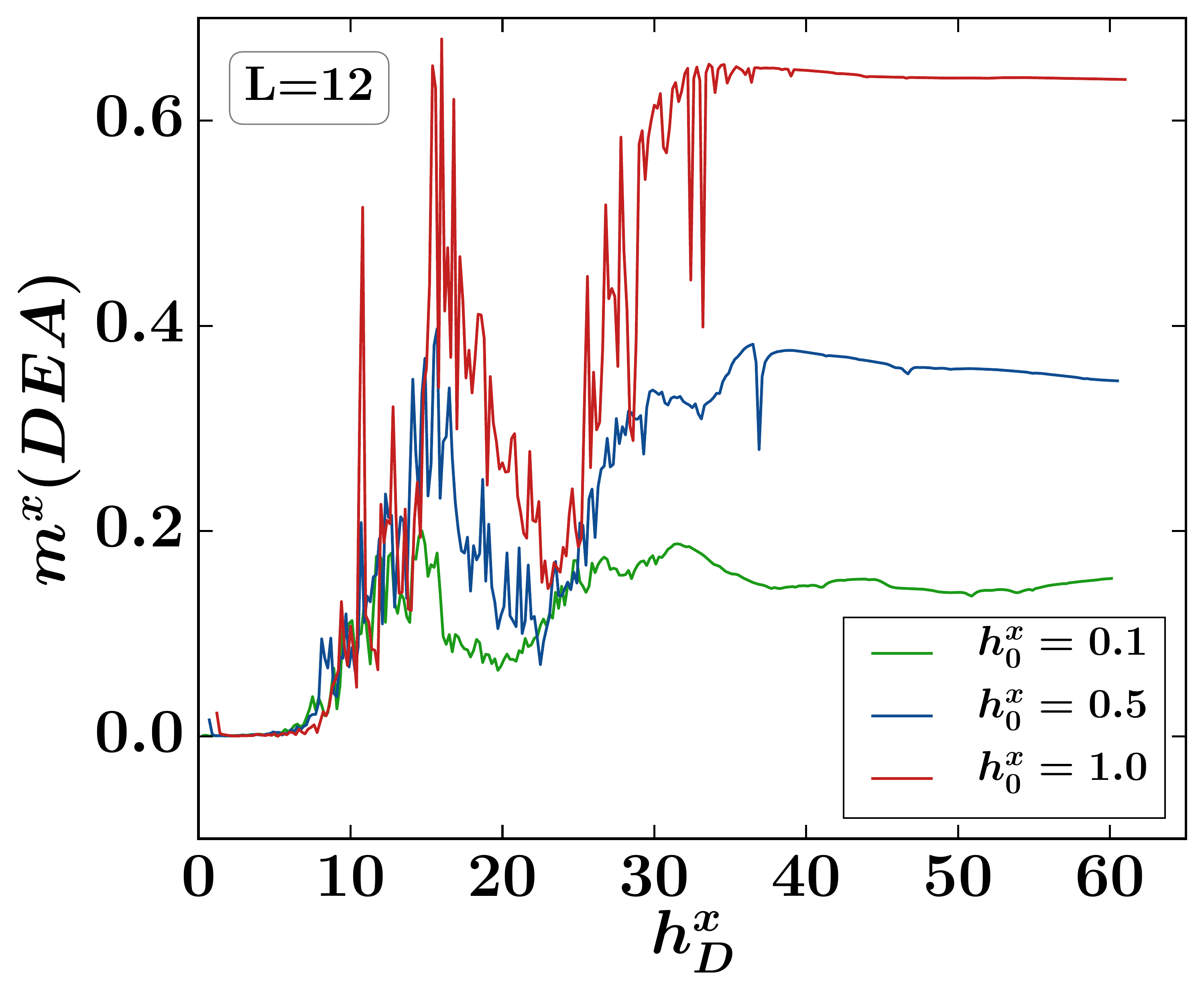}
\includegraphics[width=0.35\linewidth]{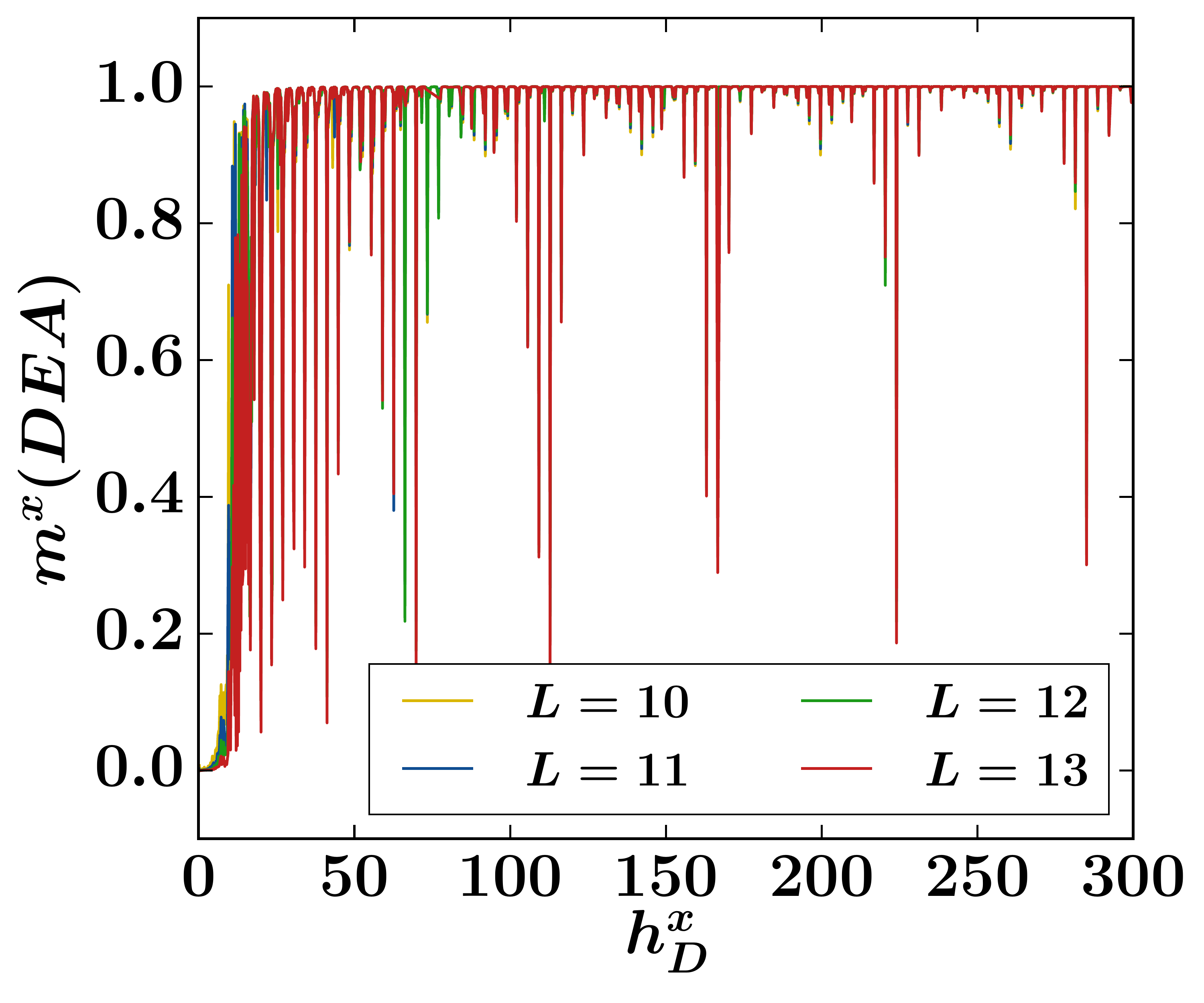}
\includegraphics[width=0.35\linewidth]{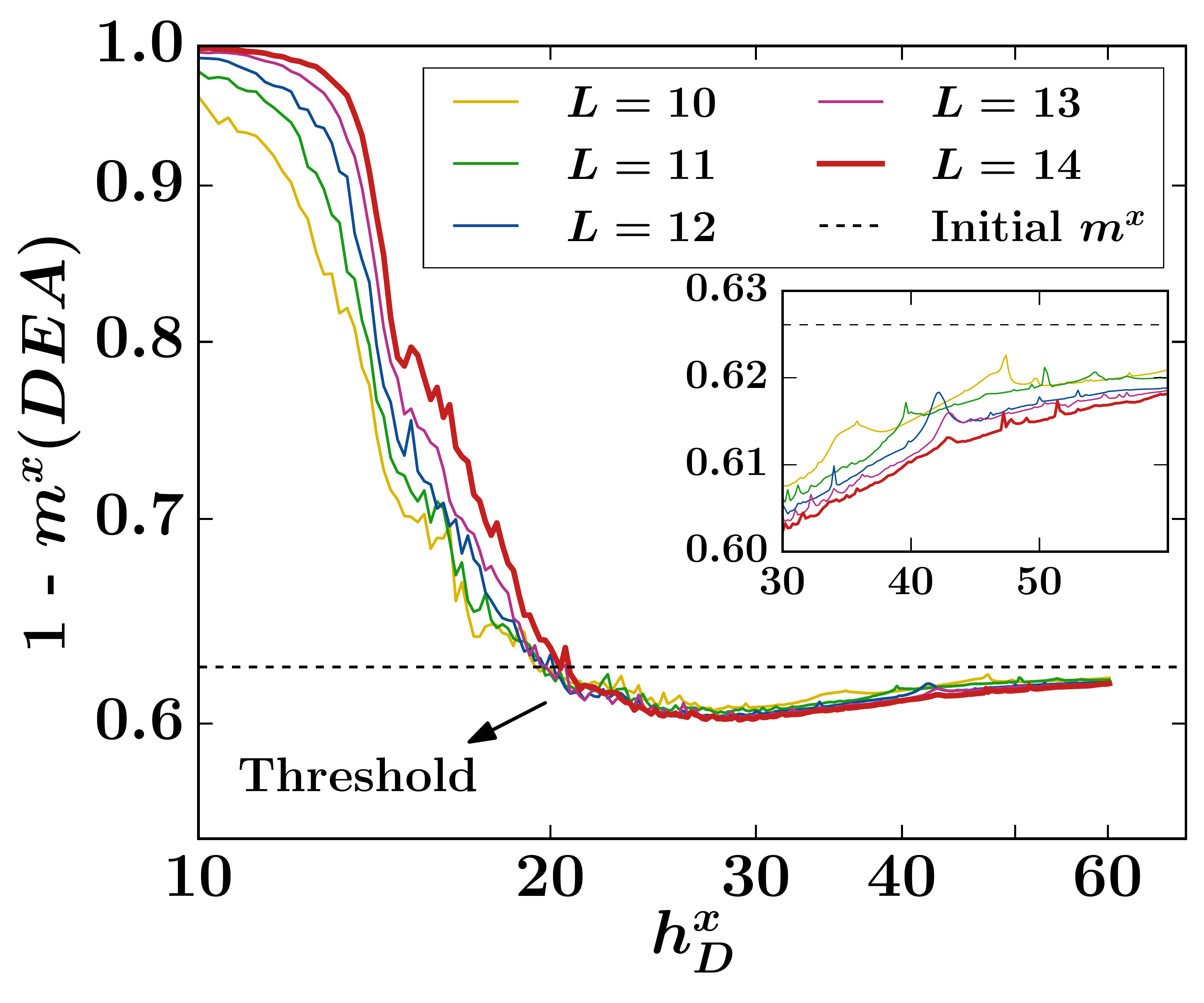}
\caption{Remnant magnetisation in various settings. 
	{\it Top left:} Dependence of a DC the transverse field $h^{z}$ which does not commute 
	with  the other, mutually commuting, terms of the model. $h^{z}$ enhances thermalization 
	(upper inset). The response approximately scales with $h_D^x/h^z$ (main panel); in particular,
	the estimated threshold ${h^x_{D}}^*$ is approximately proportional to $h^z$ (lower inset).
	{\it Top right:}  Robustness of freezing with respect to addition of DC field 
	$h_{0}^{x}$.
	{\it Bottom left:} Freezing for uneven 
	division of the total drive period. For $0\leq t < rT,$ $h_{D}^{x} = + 40,$ while for $rT\leq t < T,$
	$h_{D}^{x} = - 40$, where $r=$1/(Golden Ratio). Deep freezing minima persist to high driving
	strengths but show little size dependence. 
	{\it Bottom right:} Behaviour for initial state chosen as  the ground state of the non-integrable undriven 
	part $H_{0},$ with $h^{z}$ and $h^{x}_{0}$ chosen to created an initial state
	with a small positive polarization $m^{x}(0) \approx 0.37390.$ For large $h_D^x$, freezing increases somewhat 
	with $L.$ 
	}
	\label{Fig:Robust}
\end{figure*}

Complementary information can be gleaned by considering the correlations encoded in the non-ergodic states. 
The middle frame of Fig.~\ref{Flq_IPR} shows the magnetisation of different Floquet eigenstates, $m_i^x$,  
ordered according to their size. In the ergodic regime, these curves are featureless and 
$m_i^x$ is uniformly tiny, showing a tendency to increase with increasing drive strength. 
Deep into the nonergodic regime, large values of $m_i^x$
appear, which together form plateaux. For the largest drives $h_D^x$, the plateaux
correspond to essentially an integer number of spin flips, which indicates that 
the new basis is close to the computational basis in the $x-$direction mentioned above. As the drive is decreased,
the plateaux give way to a smooth curve, which however still makes large excursion toward $m^x=\pm1$ before 
assuming the featureless shape of the ergodic regime. 

While the fraction of Floquet states with a magnetisation above a certain value is 
thermodynamically vanishing, their entropy is
nonetheless finite, see Fig.~\ref{Flq_IPR}, middle and right frames. 
This is analogous to the case of a finite-temperature ensemble of
a magnet in a field, where a nonzero magnetisation arises as a thermodynamically vanishing fraction of 
magnetised states is preferentially populated, with their energy gain compensating for 
the entropy loss involved in 
concentrating the probability density on them. Here, the selection of the magnetised Floquet states arises 
via the  state initialisation. It is interesting to note that in this 1D system there would be no magnetization at 
any finite temperature: the observation of a finite magnetization at finite energy density is purely a 
non-equilibrium effect.

\begin{figure}[h]
\centering
\includegraphics[width=0.9\linewidth]{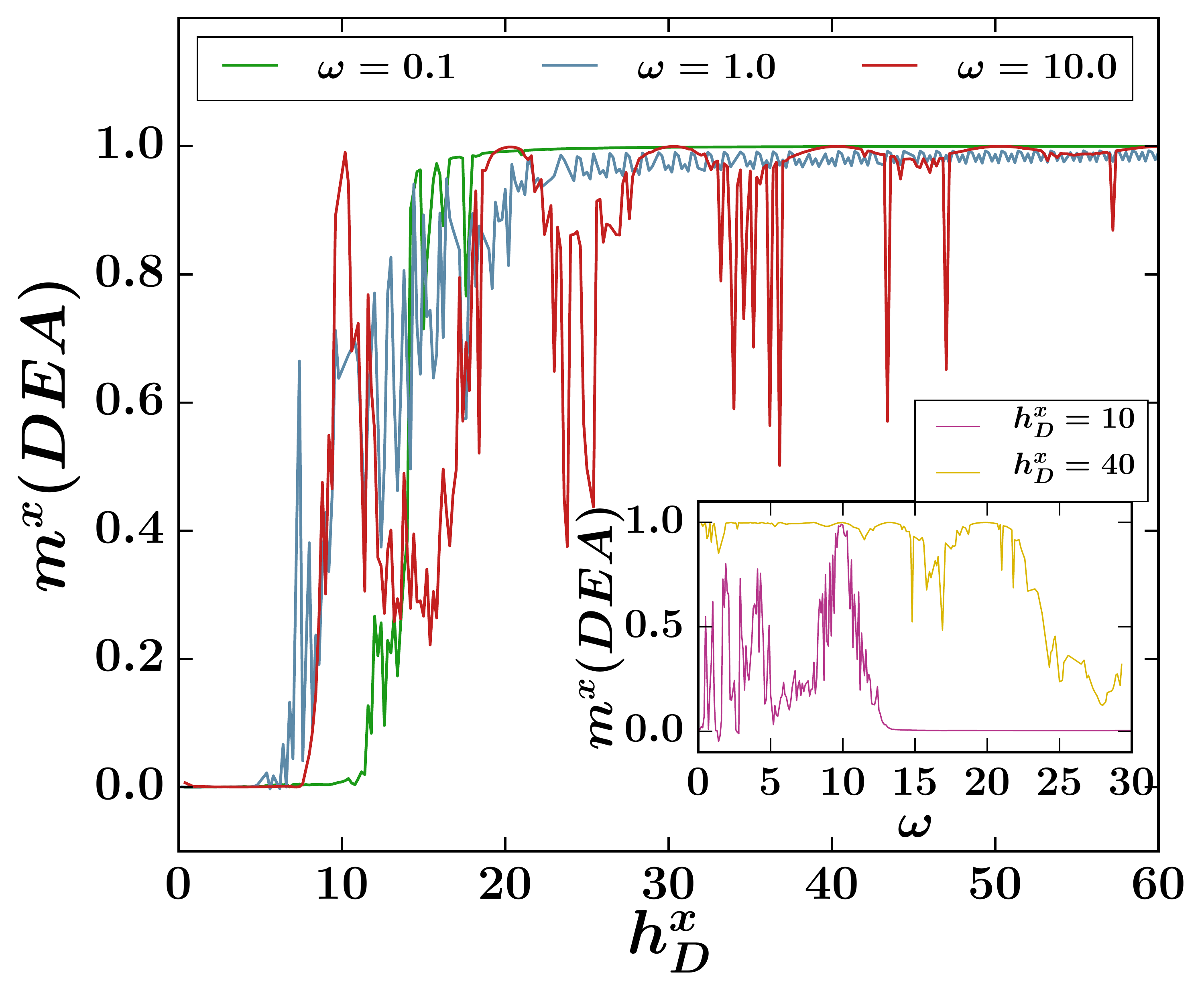}
	\caption{Dependence of the remnant magnetisation on the drive frequency $\omega$ for $L=14$ initialised
	with the ground state of $H(0)$. The basic morphology (in particular, the two regimes and the threshold) 
	remains the same over two decades in frequency. The freezing decreases as $\omega$ is increased, 
	with the threshold only varying slowly with $\omega$. The inset shows $m^x$ vs $\omega$ for 
	$h^{x}_{D} = 10$ (outside the frozen regime) and $h^{x}_{D}=40$, where the weakening of freezing with 
	increasing $\omega$ is evident.
}
\label{omega-dependence}
\end{figure}

\subsubsection{Emergence of $m^x$ as a local quasi-conserved quantity}
\comm{We next address what we believe is the central feature underpinning the non-thermalisation, namely the 
existence of a conserved quantity in the drive Hamiltonian in isolation. In our example, this is the magnetisation in the
$x$-direction, $m^x$, which persists as a quasi-conserved quantity even when the ratio of drive to static 
components of the Hamiltonian is finite.}

\comm{The middle frame of  Fig.~\ref{Flq_IPR} shows the value of $m^x$ for the different Floquet eigenstates arranged by
their size. For the strongest drives, the steps in this quantity are identical to the ones of the computational basis states in
the $x$-basis, i.e.\ the steps simply reflect the number of spins flipped.}

\comm{The static part of the Hamiltonian then mixes the states with the same value of $m^x$, which is reflected in the 
non-trivial distribution of the IPR of the Floquet states (left frame of Fig.~\ref{Flq_IPR}). The growth of the size of each 
$m^x$ sector (except for the fully polarised one) 
is in turn reflected in a decrease of the IPR.}

\comm{For lower driving strengths, $h_D^x=18$, the steps get washed out, but the range of $m^x$ continues to span practically
the full range in the interval between $-1$ and $1$. This feature disappears below the threshold, $h_D^x=10$, where the curve
flattens substantially.}

\comm{While the fraction of Floquet states with a non-zero magnetisation density vanishes with system size, these states nonetheless
have nonzero entropy, Fig.~\ref{Flq_IPR} right panel, as is the case for magnetised states of a paramagnet generally.}

\comm{The emergent quasi-conserved nature of $m^x$, along with the straightforward possibility of initialising the system 
in a magnetised state, acount for the main features of the results discussed in this work.}


\subsection{Robustness against variation of model and protocol parameters}
We first address the existence of the onset for variants of the above model. We note that so far,
no fine-tuning was necessary. The central demand was for the drive amplitude $h_D^x$ to be the largest scale,
while the other parameters of the Hamiltonian were chosen all to be in the same ballpark. 

\subsubsection{Role of non-commuting term}
First, the location of the thermalisation threshold can  be moved by varying the strength of the term in the 
static Hamiltonian $H_0$ which does not commute with the driving Hamiltonian $H_D$. Indeed, the top  left frame
of Fig.~\ref{Fig:Robust}  shows that the threshold driving field is approximately proportional to the static 
transverse field strength $h^z$.

\subsubsection{Drive shape and initial state}

\comm{Also, we ask whether the `symmetry' of having a vanishing 
mean drive of zero for symmetric pulse shapes about zero is an important ingredient. 
Fig~\ref{Fig:Robust}, top right frame, shows that the freezing is quite robust to addition of a  dc field of strength $h^{x}_{0}$.
Indeed,  the freezing actually grows with $h^{x}_{0}$.}

\comm{Next, we} consider a deviation of the drive protocol away from a time-symmetric switch in the sign of the driving term to one
where more time is spent for one sign than the other (Fig.~\ref{Fig:Robust} bottom left frame). While the latter case has
considerably more structure at high drives, in particular an apparently regular suppression of the remnant magnetisation
even above the onset threshold, the former curve basically acts as a high-magnetisation envelope of the latter. 

\comm{Further, we consider an initial state prepared as the ground-state of a many-body problem (rather than a more simply
prepared polarised state). This displays (Fig.~\ref{Fig:Robust} bottom right frame) 
all the salient features observed with the simply polarized ground state in
Fig.~\ref{fig1:Dynamics_mxDEA}, right frame.}

\subsubsection{Drive frequency}

What is particularly worth emphasizing is that the non-ergodic behaviour is {\it not} a high-frequency phenomenon. 
While such freezing also exists in the limit of a driving frequency in excess of the many-body bandwidth of the 
finite-size system, it is not even the case that the nonergodicity necessarily grows with frequency. This is 
illustrated in Fig.~\ref{omega-dependence}, where the remnant magnetisation is, if anything, more 
robust at small driving frequencies. 

\comm{This is intriguing since at  lower drive frequencies, Magnus-type 
high frequency expansions are divergent. Hence, this is an example of the breakdown of a Magnus expansion
which is not associated with unbounded heating.}

	\subsection{Finite-size behaviour}
	Our results  indicate that absence of thermalization in this 
	driven interacting system might persist even in the infinite-size limit. While there are some 
	dips of the freezing strength in the nonergodic regime complicating a sharp identification of a threshold, the 
	onset nonetheless appears to sharpen with increasing system-size. A closer view of the nonergodic 
	regime, Fig.~\ref{fig1:Dynamics_mxDEA}, middle frame top inset, shows smooth behaviour of the 
	remnant magnetisation for the largest fields; this in fact {\it grows} with increasing  system size. By contrast, for 
	weak drives, the remnant magnetisation tends to decrease with system size. This results in a
	crossing point
         as the curves for different system sizes of the deviation of the remnant magnetisation 
	from its initial value, Fig.~\ref{fig1:Dynamics_mxDEA}, right frame, thus approximately cross at
	the threshold point. While it is hard entirely to rule out a slow drift to higher fields of the threshold with
	increasing system size, 
	these observations suggest the possibility of a sharp transition at a finite threshold field in 
	the thermodynamic limit. 

	Next, and most importantly, the step-structures in the $m^x$ of the Floquet states are almost
	indistinguishable from that of the $x-$basis states for {\it all system-sizes} we investigated
	(Fig.~\ref{Flq_IPR}, middle frame). This absence of system-size dependence indicates 
	that at large values of $h_{D}^{x}$, the drive does not mix the $x-$basis states of different 
	$m^x$ values. A decrease in the fraction of Floquet states with $m^x > m_c$ with system-size is not 
	because in larger systems the Floquet states are more delocalized between different magnetization 
	sectors, but merely because the number of $x-$basis states in a given magnetization 
	sector changes with the system-size. Delocalization between different magnetization sectors is 
	suppressed {strongly} {\it for all} system-sizes at hand for $h_{D}^{x}$ above the threshold. 
	\comm{This is in keeping with the observations that on different types of initial states, 
	Fig~\ref{fig1:Dynamics_mxDEA} and Fig.\ref{Fig:Robust}, the freezing at the highest frequencies 
	does not decrease with system size.} and gives a further indication that our results are not merely 
	finite-size effects.

\section{Discussion} 
We have studied the onset of Floquet thermalisation in a driven interacting spin chain. We have found a
fairly sharp threshold for the drive strength, above which Floquet thermalisation does not take place.
The threshold value varies in different manners with parameters
like pulse shape, drive frequency, or the (non-commuting) transverse field strength, but the freezing
persists robustly under all these variations. The question of the existence of such a threshold is of 
fundamental importance, with a related issue appearing for classical dynamical systems, where the 
Kolmogorov-Arnold-Moser theorem deals with the onset of chaotic behaviour upon breaking of integrability.

       An open question is the origin, and in particular the $L$-dependence, of the dips in the frozen component
        even beyond the threshold in the $m^x$ vs $h_{D}^{x}$ plots: 
        the dips touching the $x-$axis 
	correspond to points of thermalization. While their occurrence for certain discrete 
	values of $h_{D}^{x}$ has no significant consequence, if their number diverges with
	 $L$, this may lead to a destruction of the frozen regime. For drives with pulse durations
	evenly placed about $T/2$, the dips disappear rapidly with increasing $h_{D}^{x}$. Such dips are, 
	however, observed to persist even for very strong amplitudes for the case of drive with uneven 
	division of the drive period (Fig.~\ref{Fig:Robust}, bottom left). In this case, the total drive period 
	is divided in two parts, $T/GR$ and $T(1-1/GR),$ where $GR$ is the Golden ratio. While the depth
	of the dips seems to increase with $L,$ their number and locations remain surprisingly independent
	of $L,$ which points against their proliferation. Regarding an extrapolation to the thermodynamic limit,
	we refer to our discussion at the end of the previous section.

Comparison of the magnetization and IPR of the Floquet states in the frozen regime allows one to conclude 
that the magnetization itself plays the role of a quasi-conserved quantity, which becomes exactly conserved 
in the limit of infinitely strong driving. However, the emergence of only a single conserved quantity does 
not rule out non-trivial steady states, as can be gleaned from the structure of Floquet eigenstates in the 
frozen regime: these states have definite $m^x$ values yet they are not fully localised in the $x-$basis. 
It is also interesting to note a single local conserved quantity like $m^x$ does not preclude a non-local 
$H_{eff},$ yet is sufficient to result in a non-thermal Floquet spectrum.

While our driving term in isolation is integrable, it appears that the existence of a conserved quantity is all that is
required for the existence of the frozen regime. A study of a non-integrable drive with an emergent conservation law is
therefore an obvious item for future work.  

This non-ergodicity is {\it not} a high-frequency phenomenon. Instead, it is particularly well-developed at 
lower driving frequencies, which a priori renders attempts to construct a Magnus-type high frequency expansion 
problematic. Instead, non-ergodicity is primarily associated  with strong driving. Note that for the driving term in isolation,
the instantaneous eigenvectors of the Hamiltonian are time-independent, while the instantaneous eigenvalues
change; this suggests the development of a perturbation theory controlled by the instantaneous gap, rather than a high
frequency. It would also be interesting to investigate the connections of this problem to the case of weakly driven 
interacting systems with approximate conservation laws\cite{Zala_Rosch}.

The role of emergent conservation laws may in particular 
be important for experimental studies of driven many-body systems. Indeed, a first sighting of the 
physics we have analysed here has occurred in the context of an experiment of Floquet many-body localisation 
\cite{Bordia_Knap_Bloch}, where the possibility of a finite
threshold for delocalisation was also noted for the low-disorder limit. 
The main ingredient we have identified, an emergent conservation law,  
turns out to also have been present in that situation. Analogously, for the searches of time crystals taking place at 
present, it will be interesting to investigate if emergent conservation laws do, or can, play a role there as well. 

Finally, while periodic driving is expected to heat a system and hence delocalize it, drive-induced
destructive quantum interference can produce just the opposite effect. Competition of these might 
result in unexpected freezing behaviour, as has been observed in quantum counterpart of classically chaotic systems,
namely, in the kicked rotators (see, e.g., ~\cite{Stuckman_book}). Such a suppression of heating~\cite{Haldar_Das} 
might not be impossible in a quantum many-body system 
where interactions lead to ergodicity. An absence of unbounded heating under periodic driving could be a step 
in that direction, and the availability of emergent approximate conservation laws may turn out to be a 
useful ingredient for many-body Floquet engineering. 

\section*{Acknowledgements}
The authors acknowledge  P. Bordia, A. Eckardt, V. Khemani, M. Knap and A. Polkovnikov  for discussions. 
AD and AH acknowledge the partner group program ``Spin liquids: correlations, dynamics
and disorder" between IACS and MPI-PKS, and the visitor's program of MPI-PKS for supporting visits 
to PKS during the collaboration.  RM is grateful to IACS for hospitality during the conclusion of this work.

\bibliography{Sqr_Puls_HMD}

\begin{thebibliography}{35}
\expandafter\ifx\csname natexlab\endcsname\relax\def\natexlab#1{#1}\fi
\expandafter\ifx\csname bibnamefont\endcsname\relax
  \def\bibnamefont#1{#1}\fi
\expandafter\ifx\csname bibfnamefont\endcsname\relax
  \def\bibfnamefont#1{#1}\fi
\expandafter\ifx\csname citenamefont\endcsname\relax
  \def\citenamefont#1{#1}\fi
\expandafter\ifx\csname url\endcsname\relax
  \def\url#1{\texttt{#1}}\fi
\expandafter\ifx\csname urlprefix\endcsname\relax\def\urlprefix{URL }\fi
\providecommand{\bibinfo}[2]{#2}
\providecommand{\eprint}[2][]{\url{#2}}

\bibitem[{\citenamefont{Huang}(1976)}]{Huang_Stat_Mech}
\bibinfo{author}{\bibfnamefont{K.}~\bibnamefont{Huang}},
  \emph{\bibinfo{title}{Statistical Mechanics, (2nd Ed.)}}
  (\bibinfo{publisher}{John Wiley and Sons. Inc.}, \bibinfo{year}{1976}).

\bibitem[{\citenamefont{Lazarides
  et~al.}(2014{\natexlab{a}})\citenamefont{Lazarides, Das, and
  Moessner}}]{LDM_PRE}
\bibinfo{author}{\bibfnamefont{A.}~\bibnamefont{Lazarides}},
  \bibinfo{author}{\bibfnamefont{A.}~\bibnamefont{Das}}, \bibnamefont{and}
  \bibinfo{author}{\bibfnamefont{R.}~\bibnamefont{Moessner}},
  \bibinfo{journal}{Phys. Rev. E} \textbf{\bibinfo{volume}{90}},
  \bibinfo{pages}{012110} (\bibinfo{year}{2014}{\natexlab{a}}),
  \urlprefix\url{https://link.aps.org/doi/10.1103/PhysRevE.90.012110}.

\bibitem[{\citenamefont{D'Alessio and Rigol}(2014)}]{Rigol_Infinite_T}
\bibinfo{author}{\bibfnamefont{L.}~\bibnamefont{D'Alessio}} \bibnamefont{and}
  \bibinfo{author}{\bibfnamefont{M.}~\bibnamefont{Rigol}},
  \bibinfo{journal}{Phys. Rev. X} \textbf{\bibinfo{volume}{4}},
  \bibinfo{pages}{041048} (\bibinfo{year}{2014}),
  \urlprefix\url{https://link.aps.org/doi/10.1103/PhysRevX.4.041048}.

\bibitem[{\citenamefont{Khemani et~al.}(2016)\citenamefont{Khemani, Lazarides,
  Moessner, and Sondhi}}]{KLMS}
\bibinfo{author}{\bibfnamefont{V.}~\bibnamefont{Khemani}},
  \bibinfo{author}{\bibfnamefont{A.}~\bibnamefont{Lazarides}},
  \bibinfo{author}{\bibfnamefont{R.}~\bibnamefont{Moessner}}, \bibnamefont{and}
  \bibinfo{author}{\bibfnamefont{S.~L.} \bibnamefont{Sondhi}},
  \bibinfo{journal}{Phys. Rev. Lett.} \textbf{\bibinfo{volume}{116}},
  \bibinfo{pages}{250401} (\bibinfo{year}{2016}),
  \urlprefix\url{https://link.aps.org/doi/10.1103/PhysRevLett.116.250401}.

\bibitem[{\citenamefont{Zhang~{\it et. al.}}(2017)}]{DTCions}
\bibinfo{author}{\bibfnamefont{J.}~\bibnamefont{Zhang~{\it et. al.}}},
  \bibinfo{journal}{Nature} \textbf{\bibinfo{volume}{543}},
  \bibinfo{pages}{217} (\bibinfo{year}{2017}).

\bibitem[{\citenamefont{Choi~{\it et. al.}}(2017)}]{DTCnvcentres}
\bibinfo{author}{\bibfnamefont{S.}~\bibnamefont{Choi~{\it et. al.}}},
  \bibinfo{journal}{Nature} \textbf{\bibinfo{volume}{543}},
  \bibinfo{pages}{221} (\bibinfo{year}{2017}).

\bibitem[{\citenamefont{Rovny et~al.}(2018)\citenamefont{Rovny, Blum, and
  Barrett}}]{barrett_NMR_DTC}
\bibinfo{author}{\bibfnamefont{J.}~\bibnamefont{Rovny}},
  \bibinfo{author}{\bibfnamefont{R.~L.} \bibnamefont{Blum}}, \bibnamefont{and}
  \bibinfo{author}{\bibfnamefont{S.~E.} \bibnamefont{Barrett}}
  (\bibinfo{year}{2018}), \eprint{arXiv:1802.00126v1}.

\bibitem[{\citenamefont{Das}(2010)}]{AD-DMF}
\bibinfo{author}{\bibfnamefont{A.}~\bibnamefont{Das}}, \bibinfo{journal}{Phys.
  Rev. B} \textbf{\bibinfo{volume}{82}}, \bibinfo{pages}{172402}
  (\bibinfo{year}{2010}),
  \urlprefix\url{http://link.aps.org/doi/10.1103/PhysRevB.82.172402}.

\bibitem[{\citenamefont{Hegde et~al.}(2014)\citenamefont{Hegde, Katiyar,
  Mahesh, and Das}}]{Mahesh_Freezing}
\bibinfo{author}{\bibfnamefont{S.~S.} \bibnamefont{Hegde}},
  \bibinfo{author}{\bibfnamefont{H.}~\bibnamefont{Katiyar}},
  \bibinfo{author}{\bibfnamefont{T.~S.} \bibnamefont{Mahesh}},
  \bibnamefont{and} \bibinfo{author}{\bibfnamefont{A.}~\bibnamefont{Das}},
  \bibinfo{journal}{Phys. Rev. B} \textbf{\bibinfo{volume}{90}},
  \bibinfo{pages}{174407} (\bibinfo{year}{2014}),
  \urlprefix\url{https://link.aps.org/doi/10.1103/PhysRevB.90.174407}.

\bibitem[{\citenamefont{Russomanno et~al.}(2013)\citenamefont{Russomanno,
  Silva, and Santoro}}]{Russomanno_JStatMech}
\bibinfo{author}{\bibfnamefont{A.}~\bibnamefont{Russomanno}},
  \bibinfo{author}{\bibfnamefont{A.}~\bibnamefont{Silva}}, \bibnamefont{and}
  \bibinfo{author}{\bibfnamefont{G.~E.} \bibnamefont{Santoro}},
  \bibinfo{journal}{J. Stat. Mech.} \textbf{\bibinfo{volume}{2013}},
  \bibinfo{pages}{P09012} (\bibinfo{year}{2013}).

\bibitem[{\citenamefont{Lazarides
  et~al.}(2014{\natexlab{b}})\citenamefont{Lazarides, Das, and Moessner}}]{PGE}
\bibinfo{author}{\bibfnamefont{A.}~\bibnamefont{Lazarides}},
  \bibinfo{author}{\bibfnamefont{A.}~\bibnamefont{Das}}, \bibnamefont{and}
  \bibinfo{author}{\bibfnamefont{R.}~\bibnamefont{Moessner}},
  \bibinfo{journal}{Phys. Rev. Letts.} \textbf{\bibinfo{volume}{112}},
  \bibinfo{pages}{150401} (\bibinfo{year}{2014}{\natexlab{b}}).

\bibitem[{\citenamefont{Canovi et~al.}(2016)\citenamefont{Canovi, Kollar, and
  Eckstein}}]{Eckstein_Canovi}
\bibinfo{author}{\bibfnamefont{E.}~\bibnamefont{Canovi}},
  \bibinfo{author}{\bibfnamefont{M.}~\bibnamefont{Kollar}}, \bibnamefont{and}
  \bibinfo{author}{\bibfnamefont{M.}~\bibnamefont{Eckstein}},
  \bibinfo{journal}{Phys. Rev. E} \textbf{\bibinfo{volume}{93}},
  \bibinfo{pages}{012130} (\bibinfo{year}{2016}),
  \urlprefix\url{https://link.aps.org/doi/10.1103/PhysRevE.93.012130}.

\bibitem[{\citenamefont{Abanin et~al.}(2017)\citenamefont{Abanin, De~Roeck, Ho,
  and Huveneer}}]{Dima_Floquet_Prethermalization}
\bibinfo{author}{\bibfnamefont{D.}~\bibnamefont{Abanin}},
  \bibinfo{author}{\bibfnamefont{W.}~\bibnamefont{De~Roeck}},
  \bibinfo{author}{\bibfnamefont{W.~W.} \bibnamefont{Ho}}, \bibnamefont{and}
  \bibinfo{author}{\bibfnamefont{F.}~\bibnamefont{Huveneer}},
  \bibinfo{journal}{Commun. Math. Phys.} \textbf{\bibinfo{volume}{354}},
  \bibinfo{pages}{809} (\bibinfo{year}{2017}).

\bibitem[{\citenamefont{Else et~al.}(2017)\citenamefont{Else, Bauer, and
  Nayak}}]{else_nayak_prethermal}
\bibinfo{author}{\bibfnamefont{D.~V.} \bibnamefont{Else}},
  \bibinfo{author}{\bibfnamefont{B.}~\bibnamefont{Bauer}}, \bibnamefont{and}
  \bibinfo{author}{\bibfnamefont{C.}~\bibnamefont{Nayak}},
  \bibinfo{journal}{Phys. Rev. X} \textbf{\bibinfo{volume}{7}},
  \bibinfo{pages}{011026} (\bibinfo{year}{2017}),
  \urlprefix\url{https://link.aps.org/doi/10.1103/PhysRevX.7.011026}.

\bibitem[{\citenamefont{Seetharam et~al.}(2018)\citenamefont{Seetharam, Titum,
  Kolodrubetz, and Refael}}]{gil-fss}
\bibinfo{author}{\bibfnamefont{K.}~\bibnamefont{Seetharam}},
  \bibinfo{author}{\bibfnamefont{P.}~\bibnamefont{Titum}},
  \bibinfo{author}{\bibfnamefont{M.}~\bibnamefont{Kolodrubetz}},
  \bibnamefont{and} \bibinfo{author}{\bibfnamefont{G.}~\bibnamefont{Refael}},
  \bibinfo{journal}{Phys. Rev. B} \textbf{\bibinfo{volume}{97}},
  \bibinfo{pages}{014311} (\bibinfo{year}{2018}),
  \urlprefix\url{https://link.aps.org/doi/10.1103/PhysRevB.97.014311}.

\bibitem[{\citenamefont{Mondal et~al.}(2012)\citenamefont{Mondal, Pekker, and
  Sengupta}}]{Kris-Periodic}
\bibinfo{author}{\bibfnamefont{S.}~\bibnamefont{Mondal}},
  \bibinfo{author}{\bibfnamefont{D.}~\bibnamefont{Pekker}}, \bibnamefont{and}
  \bibinfo{author}{\bibfnamefont{K.}~\bibnamefont{Sengupta}},
  \bibinfo{journal}{EPL} \textbf{\bibinfo{volume}{100}}, \bibinfo{pages}{60007}
  (\bibinfo{year}{2012}),
  \urlprefix\url{http://dx.doi.org/10.1209/0295-5075/100/60007}.

\bibitem[{\citenamefont{D'Alessio and Polkovnikov}(2013)}]{Luca_Polku}
\bibinfo{author}{\bibfnamefont{L.}~\bibnamefont{D'Alessio}} \bibnamefont{and}
  \bibinfo{author}{\bibfnamefont{A.}~\bibnamefont{Polkovnikov}},
  \bibinfo{journal}{Annals of Physics} \textbf{\bibinfo{volume}{333}},
  \bibinfo{pages}{19} (\bibinfo{year}{2013}).

\bibitem[{\citenamefont{Bukov et~al.}(2016)\citenamefont{Bukov, Heyl, Huse, and
  Polkovnikov}}]{Bukov_Polku_Huse}
\bibinfo{author}{\bibfnamefont{M.}~\bibnamefont{Bukov}},
  \bibinfo{author}{\bibfnamefont{M.}~\bibnamefont{Heyl}},
  \bibinfo{author}{\bibfnamefont{D.~A.} \bibnamefont{Huse}}, \bibnamefont{and}
  \bibinfo{author}{\bibfnamefont{A.}~\bibnamefont{Polkovnikov}},
  \bibinfo{journal}{Phys. Rev. B} \textbf{\bibinfo{volume}{93}},
  \bibinfo{pages}{155132} (\bibinfo{year}{2016}),
  \urlprefix\url{https://link.aps.org/doi/10.1103/PhysRevB.93.155132}.

\bibitem[{\citenamefont{Bukov et~al.}(2015)\citenamefont{Bukov, D'Alessio, and
  Polkovnikov}}]{Anatoli_Rev}
\bibinfo{author}{\bibfnamefont{M.}~\bibnamefont{Bukov}},
  \bibinfo{author}{\bibfnamefont{L.}~\bibnamefont{D'Alessio}},
  \bibnamefont{and}
  \bibinfo{author}{\bibfnamefont{A.}~\bibnamefont{Polkovnikov}},
  \bibinfo{journal}{Advances in Physics} \textbf{\bibinfo{volume}{64}},
  \bibinfo{pages}{139} (\bibinfo{year}{2015}).

\bibitem[{\citenamefont{Agarwala and Sen}(2017)}]{Adhip_Diptiman}
\bibinfo{author}{\bibfnamefont{A.}~\bibnamefont{Agarwala}} \bibnamefont{and}
  \bibinfo{author}{\bibfnamefont{D.}~\bibnamefont{Sen}},
  \bibinfo{journal}{Phys. Rev. B} \textbf{\bibinfo{volume}{95}},
  \bibinfo{pages}{014305} (\bibinfo{year}{2017}),
  \urlprefix\url{https://link.aps.org/doi/10.1103/PhysRevB.95.014305}.

\bibitem[{\citenamefont{Dasgupta et~al.}(2015)\citenamefont{Dasgupta,
  Bhattacharya, and Dutta}}]{Sayak_Utsa_Amit}
\bibinfo{author}{\bibfnamefont{S.}~\bibnamefont{Dasgupta}},
  \bibinfo{author}{\bibfnamefont{U.}~\bibnamefont{Bhattacharya}},
  \bibnamefont{and} \bibinfo{author}{\bibfnamefont{A.}~\bibnamefont{Dutta}},
  \bibinfo{journal}{Phys. Rev. E} \textbf{\bibinfo{volume}{91}},
  \bibinfo{pages}{052129} (\bibinfo{year}{2015}),
  \urlprefix\url{https://link.aps.org/doi/10.1103/PhysRevE.91.052129}.

\bibitem[{\citenamefont{Bordia et~al.}(2017)\citenamefont{Bordia, L\"{u}schen,
  Schneider, Knap, and Bloch}}]{Bordia_Knap_Bloch}
\bibinfo{author}{\bibfnamefont{P.}~\bibnamefont{Bordia}},
  \bibinfo{author}{\bibfnamefont{H.}~\bibnamefont{L\"{u}schen}},
  \bibinfo{author}{\bibfnamefont{U.}~\bibnamefont{Schneider}},
  \bibinfo{author}{\bibfnamefont{M.}~\bibnamefont{Knap}}, \bibnamefont{and}
  \bibinfo{author}{\bibfnamefont{I.}~\bibnamefont{Bloch}},
  \bibinfo{journal}{Nat. Phys.} \textbf{\bibinfo{volume}{13}},
  \bibinfo{pages}{460} (\bibinfo{year}{2017}).

\bibitem[{\citenamefont{Pal et~al.}(2017)\citenamefont{Pal, Nishad, S., and
  J.}}]{Sreejith_Mahesh}
\bibinfo{author}{\bibfnamefont{S.}~\bibnamefont{Pal}},
  \bibinfo{author}{\bibfnamefont{N.}~\bibnamefont{Nishad}},
  \bibinfo{author}{\bibfnamefont{M.~T.} \bibnamefont{S.}}, \bibnamefont{and}
  \bibinfo{author}{\bibfnamefont{S.~G.} \bibnamefont{J.}}
  (\bibinfo{year}{2017}), \eprint{arXiv:1708.08443}.

\bibitem[{\citenamefont{Qin and Hofstetter}(2018)}]{Qin_Hofstetter}
\bibinfo{author}{\bibfnamefont{T.}~\bibnamefont{Qin}} \bibnamefont{and}
  \bibinfo{author}{\bibfnamefont{W.}~\bibnamefont{Hofstetter}},
  \bibinfo{journal}{Phys. Rev. B} \textbf{\bibinfo{volume}{97}},
  \bibinfo{pages}{125115} (\bibinfo{year}{2018}),
  \urlprefix\url{https://link.aps.org/doi/10.1103/PhysRevB.97.125115}.

\bibitem[{\citenamefont{Prosen}(1998)}]{Prosen_prl_98}
\bibinfo{author}{\bibfnamefont{T.}~\bibnamefont{Prosen}},
  \bibinfo{journal}{Phys. Rev. Lett.} \textbf{\bibinfo{volume}{80}},
  \bibinfo{pages}{1808} (\bibinfo{year}{1998}),
  \urlprefix\url{https://link.aps.org/doi/10.1103/PhysRevLett.80.1808}.

\bibitem[{\citenamefont{Luitz et~al.}(2018)\citenamefont{Luitz, Lazarides, and
  Bar~Lev}}]{AL_DL_PRB}
\bibinfo{author}{\bibfnamefont{D.~J.} \bibnamefont{Luitz}},
  \bibinfo{author}{\bibfnamefont{A.}~\bibnamefont{Lazarides}},
  \bibnamefont{and} \bibinfo{author}{\bibfnamefont{Y.}~\bibnamefont{Bar~Lev}},
  \bibinfo{journal}{Phys. Rev. B} \textbf{\bibinfo{volume}{97}},
  \bibinfo{pages}{020303} (\bibinfo{year}{2018}),
  \urlprefix\url{https://link.aps.org/doi/10.1103/PhysRevB.97.020303}.

\bibitem[{\citenamefont{Lazarides et~al.}(2015)\citenamefont{Lazarides, Das,
  and Moessner}}]{FlqMBL1}
\bibinfo{author}{\bibfnamefont{A.}~\bibnamefont{Lazarides}},
  \bibinfo{author}{\bibfnamefont{A.}~\bibnamefont{Das}}, \bibnamefont{and}
  \bibinfo{author}{\bibfnamefont{R.}~\bibnamefont{Moessner}},
  \bibinfo{journal}{Phys. Rev. Lett.} \textbf{\bibinfo{volume}{115}},
  \bibinfo{pages}{030402} (\bibinfo{year}{2015}),
  \urlprefix\url{https://link.aps.org/doi/10.1103/PhysRevLett.115.030402}.

\bibitem[{\citenamefont{Ponte et~al.}(2015)\citenamefont{Ponte,
  Papi\ifmmode~\acute{c}\else \'{c}\fi{}, Huveneers, and Abanin}}]{FlqMBL2}
\bibinfo{author}{\bibfnamefont{P.}~\bibnamefont{Ponte}},
  \bibinfo{author}{\bibfnamefont{Z.}~\bibnamefont{Papi\ifmmode~\acute{c}\else
  \'{c}\fi{}}}, \bibinfo{author}{\bibfnamefont{F.}~\bibnamefont{Huveneers}},
  \bibnamefont{and} \bibinfo{author}{\bibfnamefont{D.~A.}
  \bibnamefont{Abanin}}, \bibinfo{journal}{Phys. Rev. Lett.}
  \textbf{\bibinfo{volume}{114}}, \bibinfo{pages}{140401}
  (\bibinfo{year}{2015}),
  \urlprefix\url{https://link.aps.org/doi/10.1103/PhysRevLett.114.140401}.

\bibitem[{\citenamefont{Rehn et~al.}(2016)\citenamefont{Rehn, Lazarides,
  Pollmann, and Moessner}}]{FlqMBL3}
\bibinfo{author}{\bibfnamefont{J.}~\bibnamefont{Rehn}},
  \bibinfo{author}{\bibfnamefont{A.}~\bibnamefont{Lazarides}},
  \bibinfo{author}{\bibfnamefont{F.}~\bibnamefont{Pollmann}}, \bibnamefont{and}
  \bibinfo{author}{\bibfnamefont{R.}~\bibnamefont{Moessner}},
  \bibinfo{journal}{Phys. Rev. B} \textbf{\bibinfo{volume}{94}},
  \bibinfo{pages}{020201} (\bibinfo{year}{2016}),
  \urlprefix\url{https://link.aps.org/doi/10.1103/PhysRevB.94.020201}.

\bibitem[{\citenamefont{Reimann}(2008)}]{Reimann}
\bibinfo{author}{\bibfnamefont{P.}~\bibnamefont{Reimann}},
  \bibinfo{journal}{Phys. Rev. Lett.} \textbf{\bibinfo{volume}{101}},
  \bibinfo{pages}{190403} (\bibinfo{year}{2008}),
  \urlprefix\url{https://link.aps.org/doi/10.1103/PhysRevLett.101.190403}.

\bibitem[{\citenamefont{Lazarides and Moessner}(2017)}]{Lazarides_RM_BDE}
\bibinfo{author}{\bibfnamefont{A.}~\bibnamefont{Lazarides}} \bibnamefont{and}
  \bibinfo{author}{\bibfnamefont{R.}~\bibnamefont{Moessner}},
  \bibinfo{journal}{Phys. Rev. B} \textbf{\bibinfo{volume}{95}},
  \bibinfo{pages}{195135} (\bibinfo{year}{2017}),
  \urlprefix\url{https://link.aps.org/doi/10.1103/PhysRevB.95.195135}.

\bibitem[{\citenamefont{Rigol et~al.}(2016)\citenamefont{Rigol, Dunjko, and
  Olshanii}}]{Rigol_Nature}
\bibinfo{author}{\bibfnamefont{M.}~\bibnamefont{Rigol}},
  \bibinfo{author}{\bibfnamefont{V.}~\bibnamefont{Dunjko}}, \bibnamefont{and}
  \bibinfo{author}{\bibfnamefont{M.}~\bibnamefont{Olshanii}},
  \bibinfo{journal}{Nature} \textbf{\bibinfo{volume}{452}},
  \bibinfo{pages}{854} (\bibinfo{year}{2016}),
  \urlprefix\url{http://dx.doi.org/10.1038/nature06838}.

\bibitem[{\citenamefont{Lenar\ifmmode \check{c}\else
  \v{c}\fi{}i\ifmmode~\check{c}\else \v{c}\fi{}
  et~al.}(2018)\citenamefont{Lenar\ifmmode \check{c}\else
  \v{c}\fi{}i\ifmmode~\check{c}\else \v{c}\fi{}, Lange, and
  Rosch}}]{Zala_Rosch}
\bibinfo{author}{\bibfnamefont{Z.}~\bibnamefont{Lenar\ifmmode \check{c}\else
  \v{c}\fi{}i\ifmmode~\check{c}\else \v{c}\fi{}}},
  \bibinfo{author}{\bibfnamefont{F.}~\bibnamefont{Lange}}, \bibnamefont{and}
  \bibinfo{author}{\bibfnamefont{A.}~\bibnamefont{Rosch}},
  \bibinfo{journal}{Phys. Rev. B} \textbf{\bibinfo{volume}{97}},
  \bibinfo{pages}{024302} (\bibinfo{year}{2018}),
  \urlprefix\url{https://link.aps.org/doi/10.1103/PhysRevB.97.024302}.

\bibitem[{\citenamefont{St\"{o}ckmann}(1999)}]{Stuckman_book}
\bibinfo{author}{\bibfnamefont{H.-J.} \bibnamefont{St\"{o}ckmann}},
  \emph{\bibinfo{title}{Quantum Chaos: An Introduction}}
  (\bibinfo{publisher}{Cambridge Univ. Press}, \bibinfo{year}{1999}).

\bibitem[{\citenamefont{Haldar and Das}(2017)}]{Haldar_Das}
\bibinfo{author}{\bibfnamefont{A.}~\bibnamefont{Haldar}} \bibnamefont{and}
  \bibinfo{author}{\bibfnamefont{A.}~\bibnamefont{Das}}, \bibinfo{journal}{Ann.
  der Phys.} \textbf{\bibinfo{volume}{529}}, \bibinfo{pages}{1600333}
  (\bibinfo{year}{2017}).

\end{thebibliography}

\newpage

\section{Supplemental Material}

\subsection{Ergodicity of $H_{0}$}
\begin{figure}[H]
\centering
\includegraphics[width=0.45\linewidth]{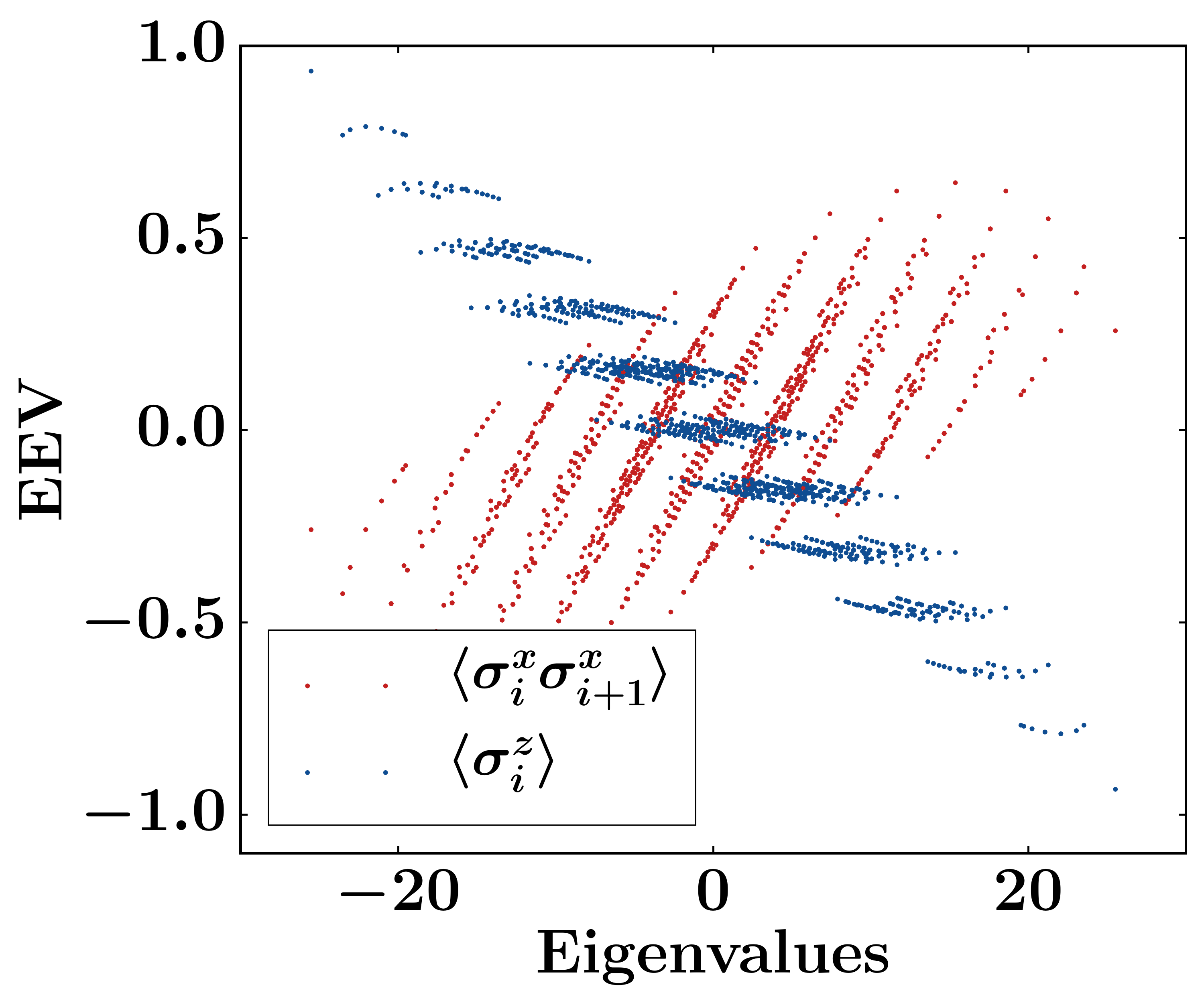}
\includegraphics[width=0.45\linewidth]{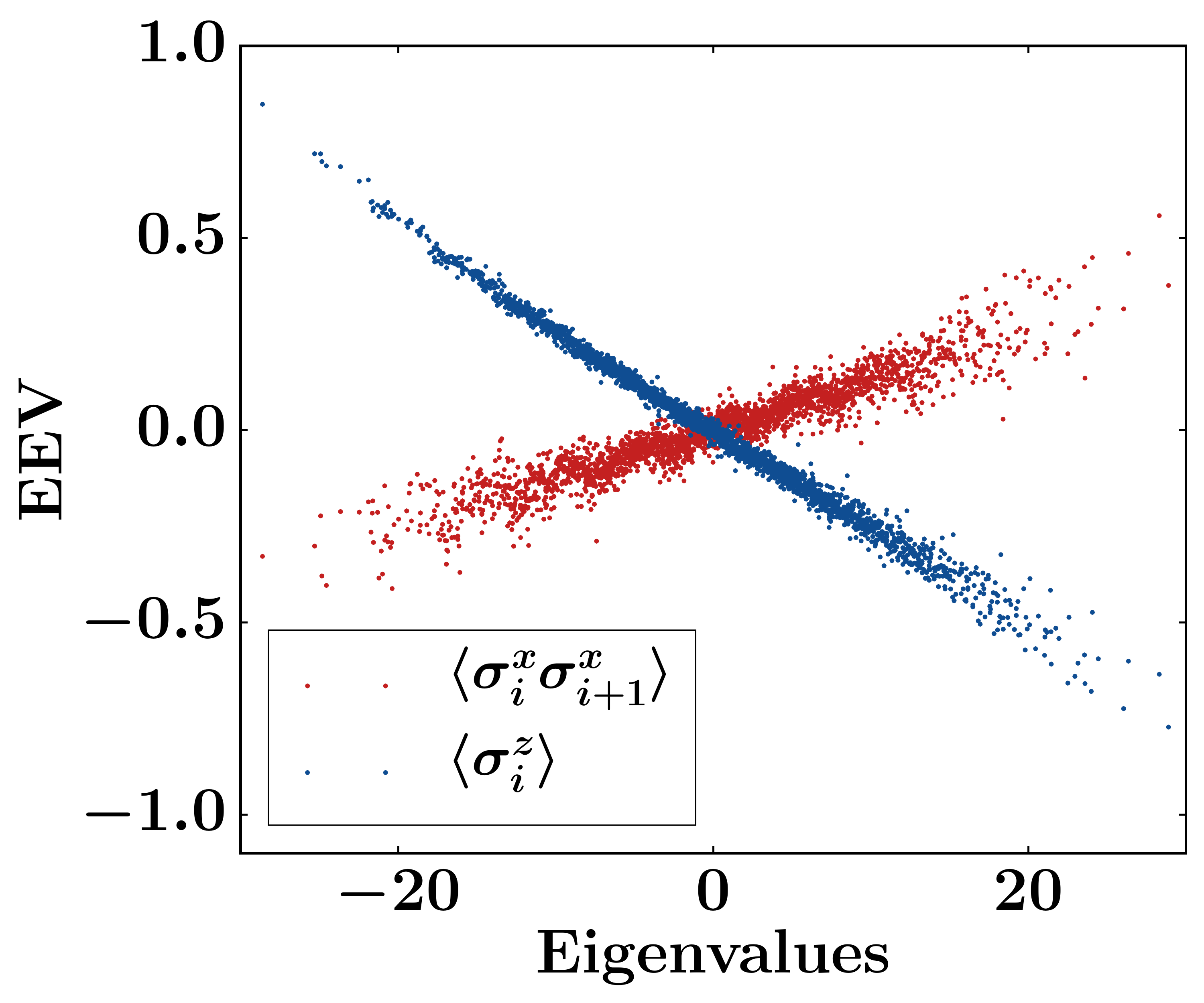}
\caption{
	Left frame shows expectation values of different local operators over the eigenstates (EEV) of the integrable 
	transverse field Ising chain ($\kappa = h^{x} = 0$) against the eigenvalues of those states $(L=12)$. The non-smooth behaviour 
	arises due to the existence of different non-mixing (block-diagonal) sectors. Right frame shows that introduction of the interaction and longitudinal
	field breaks the integrability and the EEV's become smooth, 
	confirming that the undriven Hamiltonian $H_{0}$ is quite generic and satisfies the eigenstate 
	thermalization hypothesis. The paremeters $\kappa = 0.7$ and $h^{x} = 1$ correspond to those in the main text.
}
\label{Ergodic_H0}
\end{figure}

We first demonstrate the ergodicity of the undriven Hamiltonian $H_{0}$ for the parameter values we have used in the main text.
This is displayed in Fig.~\ref{Ergodic_H0}.

\subsection{Freezing and Thermalization via Level Statistics}
Here we demonstrate the quasi-energy level repulsion and its absence for drive strengths above and below the threshold respectively.  
We plot the quasi-energy (folded to the first Brillouin zone) gap ratio defined by
$$r = \frac{min\{\delta_{n},\delta_{n-1}\}}{max\{\delta_{n},\delta_{n-1}\}},$$ 
\noindent where,
$\delta_{n} = \mu_{n+1}-\mu_{n},$ $\mu_{n}$ being the $n$-th eigenvalue of $H_{eff}$, after folding them into the first Brillouin zone [$-\pi,\pi$]. 
\subsection{Quasi-energy Gap-ratio Statistics below and above the threshold:}
\begin{figure}[H]
\centering
\includegraphics[width=0.65\linewidth]{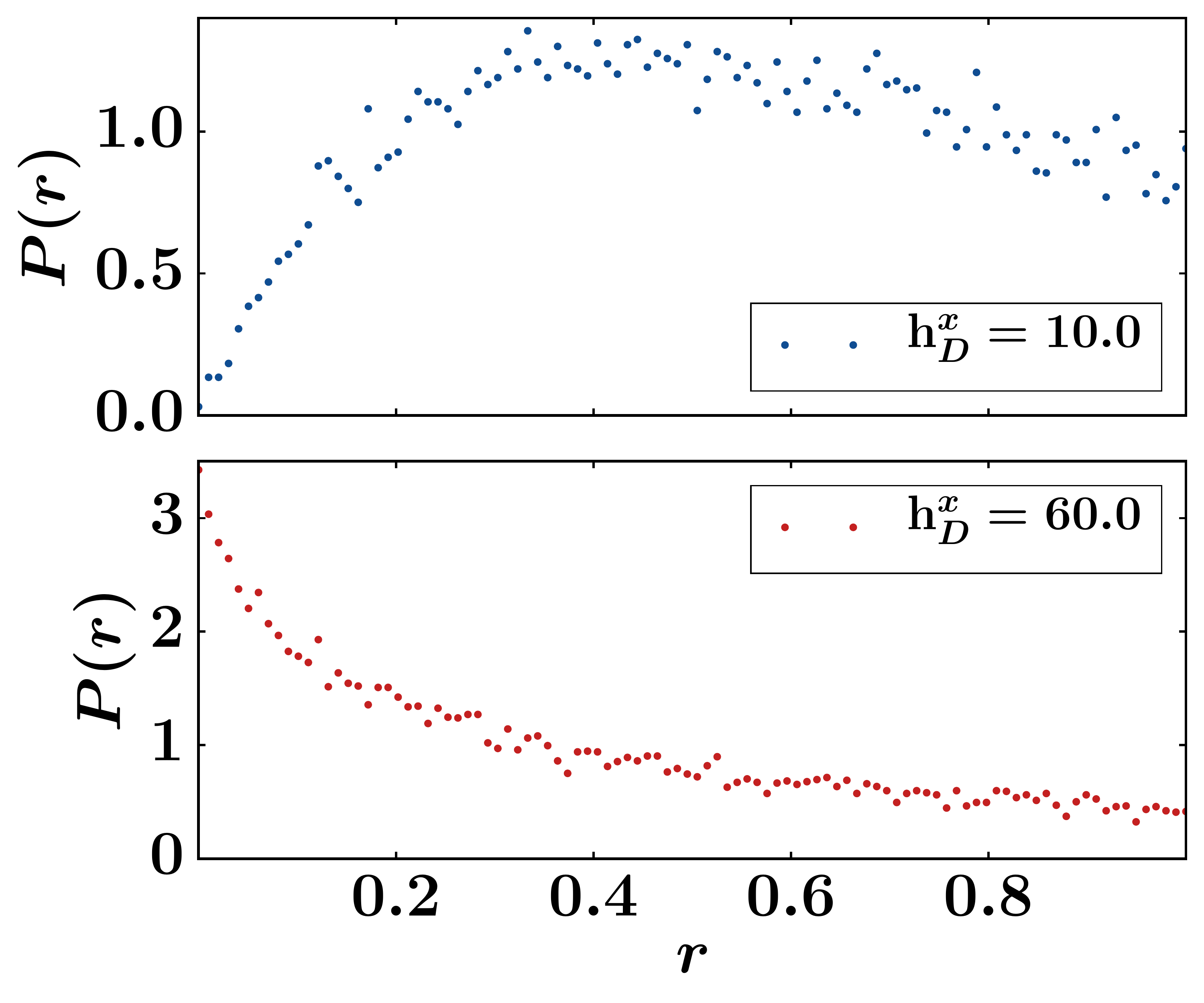}
\caption{Gap-ratio statistics $r$ for 
	$h_{D}^{x} = 10$ (below the threshold) and $60$ (far above the threshold) respectively,  
	showing presence and absence of quasi-energy level repulsion below and above the threshold respectively.	
}
\label{Gap_Stat}
\end{figure}

%
\subsection{Threshold Phenomenon with Random Polarized Initial States}
\begin{figure}[H]
\centering
\includegraphics[width=0.65\columnwidth]{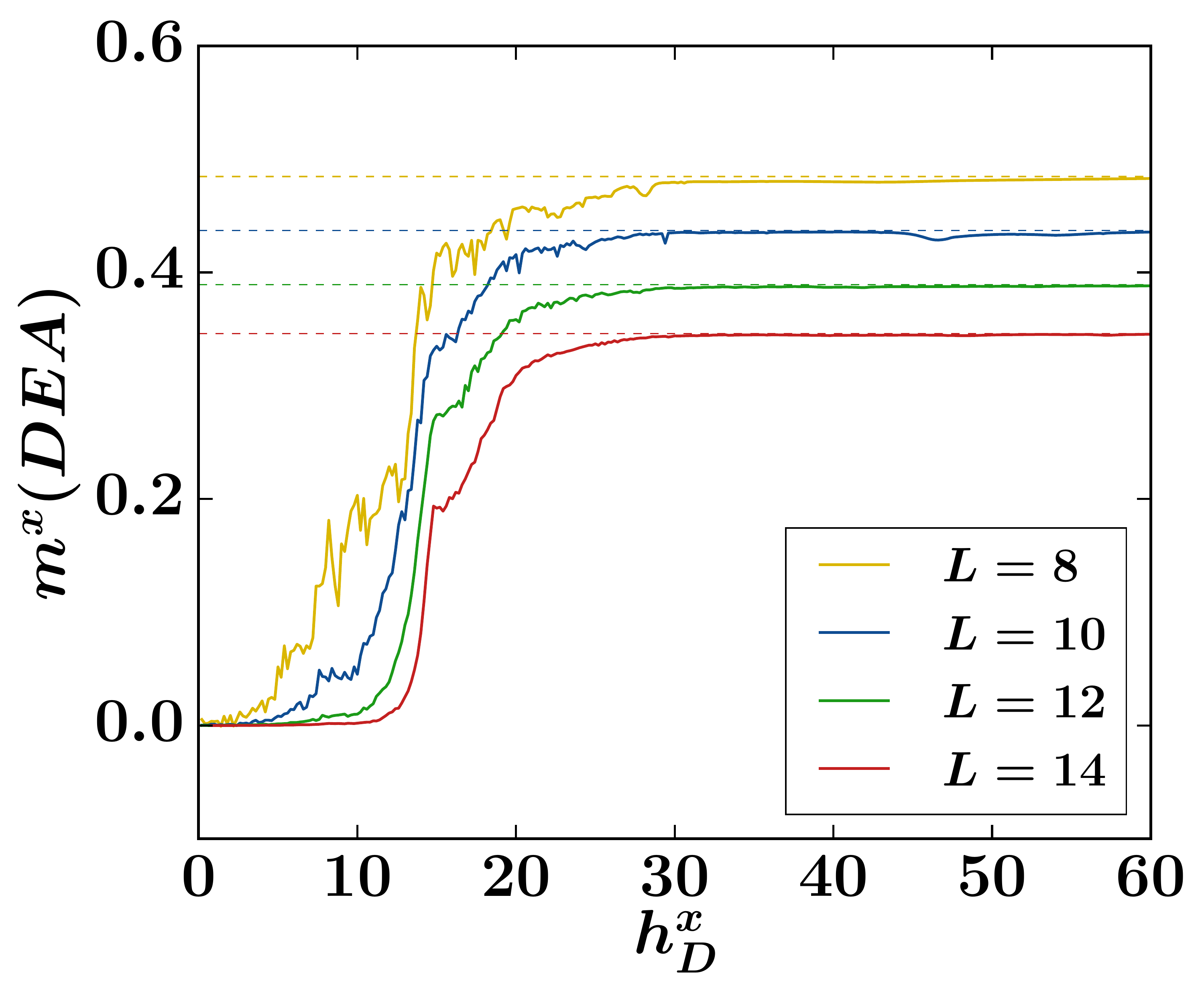}
\caption{
	Freezing for initially polarized but otherwise 
	randomised states under longitudinal drive (for the same set of parameter as in the main text). These 
	initial states are given as  $|\psi(0)\rangle = \sum_{i=1}^{2^{L}} c_{i} |i_{x}\rangle$, 
	where $|i_{x}\rangle$ is the $i-$th eigenstate in thecomputational basis 
	states in $x$-direction, by choosing $Re[c_{i}]$ and $Im[c_{i}]$ from a uniform 
	distribution between -1 and +1, multiplying them by $e^{\beta m_{i}^{x}},$ where $\beta > 0$, where  
	$m_{x}^{i}$ is the longitudinal magnetization of $|i_{x}\rangle$, and finally normalizing the state. 
	Results are shown for different $L$ values, for random instances generated with $\beta=2.5$ 
	(for which the initial magnetization decreases with $L$.) The system freezes 
	for large values of the drive field $h_{D}^{x}$ for all accessed system-sizes we could access, with
	the remanent magnetisation very close to the diagonal ensemble
	average (DEA) of the longitudinal magnetization $m^x$.  
}
\label{Flq_Eig}
\end{figure}

Fig.~\ref{Flq_Eig} shows the remanent magnetisation for a class of random polarised initial states. 

\section{$L-$dependence of the Floquet state average magnetization}
\begin{figure}[H]
\centering
\includegraphics[width=0.65\columnwidth]{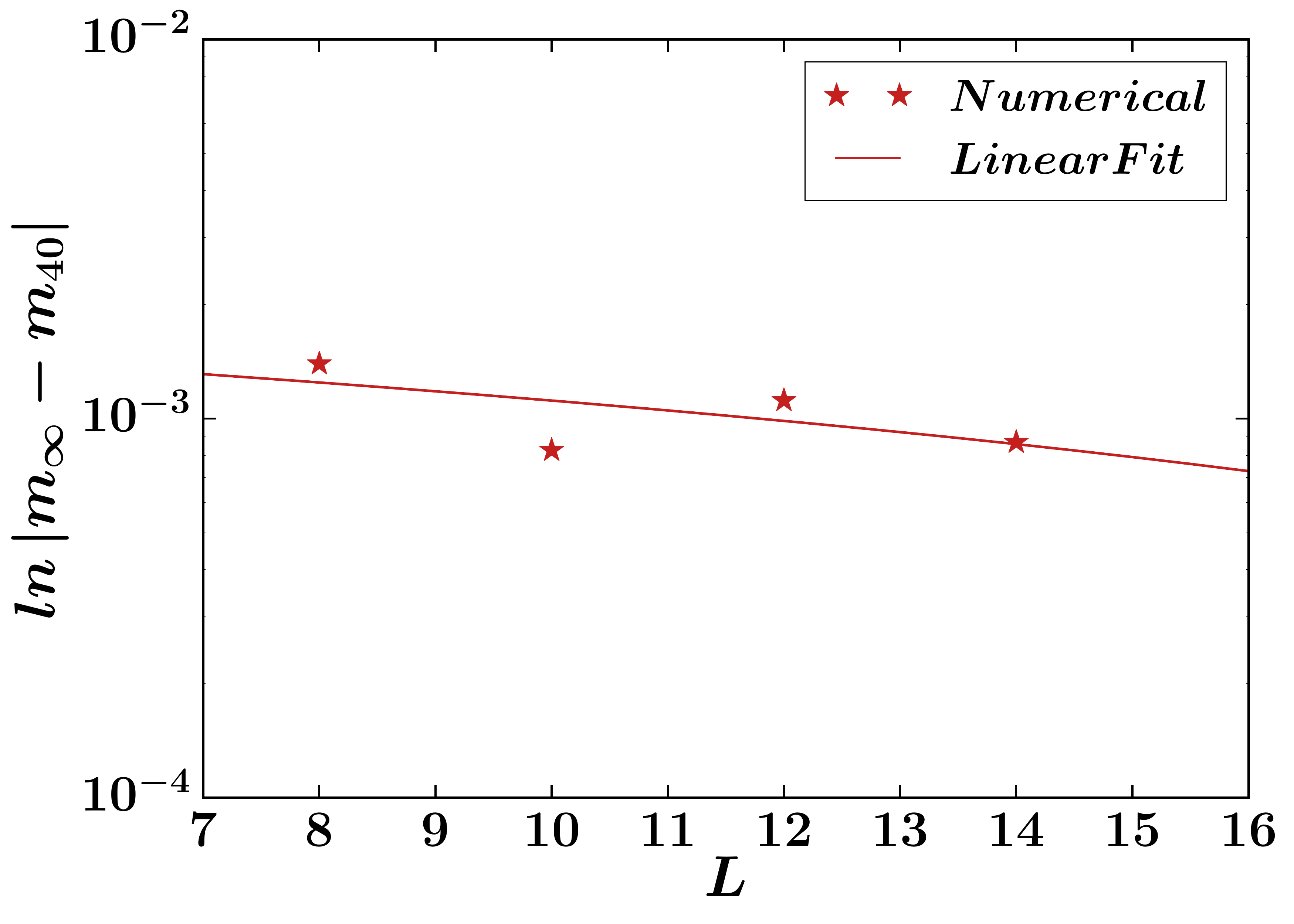}
\caption{$L-$dependence of the $log$ of the deviation of the average of the magnetization of the Floquet states from
the corresponding $x-$basis states
}
\label{mx_Diff}
\end{figure}

Fig. 2 (middle panel) of the main text shows that the step-structures in the $m^x$ of the Floquet states are almost
	indistinguishable from that of the $x-$basis states for all $L$ we investigated. 
	Here we consider the following average quantity to show on a finer scale, that this difference
	systematically decreases with $L.$ We order both the $x-$basis states and the Floquet states in order 
	(decreasing, say) of their $m^x$ values (eigenvalues and expectation values in respectively). Let 
	$m^{x}_{h_{D}^{x}}(i)$ and $m^{x}_{\infty}(i)$ denotes the $m^x$ values of the $i$-th state thus ordered in respective basis. 
	Now we compute the difference $|m_{\infty} - m_{h_{D}^{x}}|$, where 
	$m_{\infty} = \frac{1}{2^L}\sum_{i}m^{x}_{\infty}(i)$ and $m_{h_{D}^{x}} = 
	\frac{1}{2^L}\sum_{i}m^{x}_{h_{D}^{x}}(i).$ This provides a measure of the accuracy with which $m^x$ is 
	conserved (this difference vanishes if $m^x$ is exactly conserved, since in that case each Floquet state
	corresponds to an exact eigenstate of $m^x$). In Fig.~\ref{mx_Diff} we show that this difference is
	tiny, and seems, if anything, to decrease with increasing $L.$

\end{document}